\documentclass[conference]{IEEEtran}
\IEEEoverridecommandlockouts

\usepackage{booktabs}
\usepackage{makecell}
\usepackage{siunitx}

\usepackage{cite}
\usepackage{amsmath,amssymb,amsfonts}
\usepackage{graphicx}
\usepackage{textcomp}
\usepackage{xcolor}
\usepackage{verbatim}
\usepackage{placeins}
\usepackage{subcaption}
\usepackage{algorithm}
\usepackage[noend]{algpseudocode}
\floatname{algorithm}{Procedure}
\usepackage{tikz}
\usepackage{float}
\usepackage[utf8]{inputenc}
\usepackage[T1]{fontenc}
\usepackage{listings}

\newcommand{\timeout}{\textsuperscript{\textcolor{red!70!black}{\dag}}}
\newcommand{\incorrect}{\textsuperscript{\textcolor{blue!70!black}{\ddag}}}
\newcommand{\failed}{\textsuperscript{\textcolor{orange!80!black}{*}}}
\usetikzlibrary{arrows.meta, positioning, shapes, calc}
\algrenewcommand\algorithmicindent{1.2em} %

\def\BibTeX{{\rm B\kern-.05em{\sc i\kern-.025em b}\kern-.08em
    T\kern-.1667em\lower.7ex\hbox{E}\kern-.125emX}}
\makeatletter
\newcommand{\linebreakand}{%
  \end{@IEEEauthorhalign}
  \hfill\mbox{}\par
  \mbox{}\hfill\begin{@IEEEauthorhalign}
}
\makeatother

\begin{document}

\title{Distributed Datalog at Scale with Worst-Case Optimal Joins}

\author{
\IEEEauthorblockN{Kunting Qi}
\IEEEauthorblockA{\textit{University of Illinois Chicago}\\
Chicago, IL, USA\\
kqi@uic.edu}
\and
\IEEEauthorblockN{Yihao Sun}
\IEEEauthorblockA{\textit{Syracuse University}\\
Syracuse, NY, USA\\
ysun67@syr.edu}
\and
\IEEEauthorblockN{Kristopher Micinski}
\IEEEauthorblockA{\textit{Syracuse University}\\
Syracuse, NY, USA\\
kkmicins@syr.edu}
\linebreakand
\IEEEauthorblockN{Thomas Gilray}
\IEEEauthorblockA{\textit{Washington State University}\\
Pullman, WA, USA\\
thomas.gilray@wsu.edu}
\and
\IEEEauthorblockN{Sidharth Kumar}
\IEEEauthorblockA{\textit{University of Illinois Chicago}\\
Chicago, IL, USA\\
sidharth@uic.edu}
}

\maketitle

\begin{abstract}

Scalable logic-based analytics are frequently dominated by cyclic multi-way joins over large relational datasets. State-of-the-art systems evaluate these joins through chains of binary joins, incurring substantial overhead from large intermediate relations that must be materialized and redistributed across the network. Worst-Case Optimal Joins (WCOJ) offer a provably superior alternative by evaluating joins one variable at a time, eliminating intermediate blowup. However, existing WCOJ implementations target either shared-memory machines or one-shot, non-recursive queries on dataflow and big-data platforms, leaving WCOJ-based recursive evaluation on distributed-memory HPC systems an open problem. We present, to our knowledge, the first distributed-memory Datalog engine that uses WCOJ for recursive fixpoint evaluation. Our design builds on HyperCube partitioning, which hash-partitions relations along variable-aligned grid dimensions and replicates them along missing dimensions, and extends it from a single shuffle to iterative execution: in every semi-na\"ive iteration, only newly derived tuples are repartitioned and re-replicated, using MPI all-to-all collectives and per-dimension all-gathers over subcube communicators, while the same layout is reused across iterations so that all WCOJ intersections remain rank-local. We evaluate our system on the Aurora and Polaris leadership-class supercomputers at Argonne National Laboratory using graph mining and program analysis benchmarks, demonstrating strong scalability, outperforming state-of-the-art Datalog engines on most benchmarks, and characterizing how partition dimensionality trades runtime against memory feasibility.

\end{abstract}

\section{Introduction}

Large-scale relational reasoning underpins a broad class of high-impact workloads, from social network analysis and knowledge graph inference to cybersecurity threat detection and compiler pointer analysis~\cite{seo2013socialite, nenov2015rdfox, smaragdakis2015pointer, flores2020datalog}. At the core of these workloads are recursive conjunctive queries that iteratively derive new facts until a fixpoint is reached, placing sustained pressure on both memory capacity and compute throughput. As input datasets scale to billions of tuples, single-node systems quickly saturate available memory and stall on compute bottlenecks, making distributed execution across shared-nothing clusters a practical necessity. The central systems challenge is therefore to scale recursive relational query evaluation to high-performance clusters without incurring prohibitive data movement and inter-node communication costs.

These analytical workloads are naturally expressed in Datalog~\cite{abiteboul1995foundations}, a declarative logic programming language well suited to fixpoint computation over relational data. We illustrate Datalog through two representative analytics queries: the \textit{Triangle} query from graph mining and the \textit{Peer Influence} query from social network analysis. The Triangle query, $\texttt{Triangle}(x,y,z) \leftarrow \texttt{R}(x,y),\ \texttt{S}(y,z),\ \texttt{T}(z,x)$, is a non-recursive cyclic multi-way join over three relations. The \textit{Peer Influence} query, by contrast, is recursive, with the head relation feeding back into the body, and models how influence propagates through densely connected communities within a network: $\texttt{Influenced}(y) \leftarrow \texttt{Influenced}(x),\ \texttt{Edge}(x,y),\ \texttt{Edge}(x,z),\ \texttt{Edge}(z,y)$. This rule states that a vertex $y$ becomes influenced if it is adjacent to an already-influenced vertex $x$ and the two share a common neighbor $z$. Evaluating this rule requires finding such triangles by joining three occurrences of the \texttt{Edge} relation together with the \texttt{Influenced} relation, which forms a cyclic multi-way join.

\begin{figure}[t]
    \centering
    \includegraphics[width=1.0\linewidth]{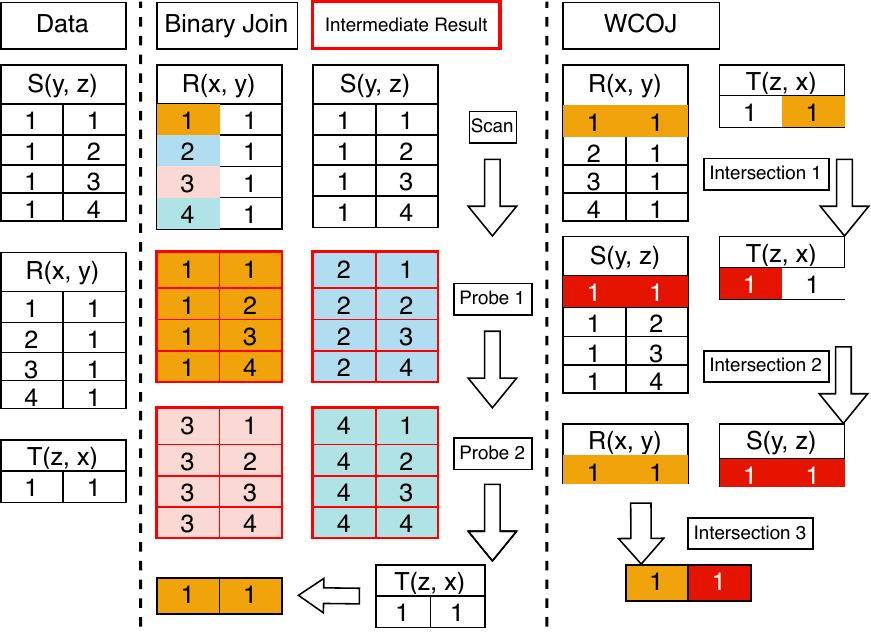}
    \caption{Comparison of join strategies for a triangle query. Binary joins (center) over-materialize intermediate results, whereas WCOJ (right) eliminates unproductive intermediates through aggressive pruning.}
    \vspace{-0.3cm}
    \label{fig:binary-vs-wcoj}
\end{figure}

Cyclic multi-way queries such as these are notoriously difficult for traditional distributed Datalog systems, which rely on binary joins. Evaluating a triangle two relations at a time forces the system to first compute all open two-hop paths. In dense graphs, these intermediate results explode, often growing asymptotically larger than both the final query output and the underlying graph itself (see Figure~\ref{fig:binary-vs-wcoj}). In a parallel setting, this intermediate blowup is costly, as the system must fully materialize, repartition, and redistribute oversized intermediate relations across the network at each evaluation step. The problem is further exacerbated in recursive settings, where such expensive two-hop joins execute repeatedly across iterations, creating a persistent communication bottleneck that fundamentally limits cluster scalability.

Worst-Case Optimal Joins (WCOJ)~\cite{10.14778/3407790.3407797, ngo2018worst} address the intermediate explosion problem with provable guarantees. Rather than decomposing a multi-way join into a sequence of pairwise joins that materialize large intermediate relations, WCOJ evaluates the join one variable at a time through simultaneous set intersection across all participating relations. At each step, only tuples that are consistent with all join constraints are extended, and candidates that cannot participate in any final output tuple are pruned immediately, before they are ever materialized (see Figure~\ref{fig:binary-vs-wcoj}). This attribute-at-a-time intersection ensures that no intermediate result ever exceeds the worst-case optimal bound established by the AGM inequality~\cite{atserias2013size}, structurally limiting intermediate data to its theoretical minimum. As a consequence, the massive shuffles and bandwidth bottlenecks incurred by repeated binary join materialization are eliminated, making WCOJ well-suited for communication-constrained distributed environments.

However, scaling WCOJ-based Datalog evaluation to a parallel system raises two challenges: \emph{how to partition the workload effectively across all processes}, and \emph{how to support fixpoint iteration for recursive queries efficiently}. Workload partitioning is inherently more complex in WCOJ than in binary join execution. Unlike binary joins, where partitioning decisions can be made independently for each relation pair, WCOJ must reconcile the partitioning requirements of all intersection variables simultaneously. For example, evaluating the \textit{Peer Influence} query requires intersecting the $x$ variable across every \texttt{Edge} and \texttt{Influenced} clause at once; a relation that does not carry a given variable cannot be partitioned on it, yet must still participate in the intersection, forcing either replication or remote data fetches. Efficient distributed WCOJ therefore demands a partition-aware execution model that jointly optimizes across all intersection variables, minimizes communication volume, and distributes the computational workload evenly across nodes. Equally important is support for recursive fixpoint iteration, where newly derived tuples from one iteration must be materialized and redistributed for use in the next, a step that again requires coordinated data movement across processes.

We address the workload partitioning challenge by developing a parallel WCOJ framework grounded in multidimensional hypercube partitioning~\cite{afrati2010optimizing,beame2017communication} (Section~\ref{sec:par-wcoj}). The core insight is that join variables naturally define dimensions of a logical process grid, and the number of variables selected for partitioning determines the dimensionality of that grid. With a single-variable partitioning scheme, the process space is divided along one dimension; relations carrying that variable are hash-partitioned across processes, while all other relations are fully replicated. In a two-variable scheme, processes are arranged in a 2D grid; relations carrying both variables are fully partitioned across the entire grid, relations carrying only one variable are partially partitioned along the corresponding dimension and require partial replication along the other, and relations carrying neither variable must be fully replicated. This pattern generalizes naturally to higher dimensions, with each additional variable typically reducing the replication of relations that carry it, at the cost of increased communication to coordinate the finer-grained partitioning. In this work, we demonstrate and evaluate partitioning schemes of up to three variables, corresponding to a 3D process grid. The choice of dimensionality exposes a tunable trade-off between replication overhead and communication volume, allowing the partitioning strategy to be adapted to the query structure and deployment scale (Section~\ref{1d2d3d}).

To support iterative WCOJ execution, we adopt semi-na\"ive evaluation (Section~\ref{iter_wcoj_par}), a standard database technique that restricts each iteration's input to the tuples newly derived in the previous iteration, avoiding redundant recomputation. Recursion is where our setting departs from prior distributed WCOJ systems~\cite{chu2015theory,ammar2018distributed,zhang2021fast}, which evaluate a single conjunctive query after one round of data shuffling. In recursive evaluation, the newly derived tuples of every iteration must be materialized, repartitioned, and re-replicated to the processes that need them in the next iteration, which in general differ from the processes that generated them. We therefore reuse the same hypercube layout across all iterations and redistribute only the newly derived tuples, using MPI all-to-all collectives followed by per-dimension all-gathers over subcube communicators, so that every WCOJ intersection remains rank-local in every iteration. In summary, this paper makes the following contributions:
\begin{enumerate}
    \item We present, to our knowledge, the first distributed-memory Datalog engine that uses WCOJ for recursive semi-na\"ive fixpoint evaluation.
    \item We extend HyperCube partitioning~\cite{afrati2010optimizing,beame2017communication} from a single shuffle to iterative execution: each iteration repartitions and re-replicates only the newly derived tuples, using \texttt{MPI\_Alltoallv} and per-dimension \texttt{MPI\_Allgatherv} over subcube communicators, and reuses the same layout across iterations so that WCOJ intersections never require remote accesses.
    \item We evaluate the system on Polaris and Aurora, demonstrating strong scaling up to 2,048 processes and competitive performance against prior single-node and distributed Datalog systems, and we characterize how partition dimensionality trades runtime against memory feasibility.
\end{enumerate}

Across the evaluated workloads, our system demonstrates strong scalability, achieving up to 14.3$\times$ speedup on Polaris and 13.4$\times$ on Aurora when scaling from 32 to 2,048 processes. On a single node, it is faster than established single-node engines on most of our recursive and cyclic query benchmarks, and it outperforms existing distributed Datalog engines on transitive closure. The remainder of this paper is organized as follows. Section~\ref{sec:background} reviews the necessary background. Section~\ref{sec:par-wcoj} presents the workload partitioning strategy of our distributed WCOJ implementation, and Section~\ref{iter_wcoj_par} describes how we extend it to the iterative setting. Section~\ref{evaluation} reports scalability results on the Polaris and Aurora supercomputers and compares our system with existing engines. Section~\ref{sec:related} discusses related work, and Section~\ref{sec:conclusion} concludes with directions for future work.

\section{Background}
\label{sec:background}

Datalog is a declarative logic programming language with historical roots as a database query language~\cite{abiteboul1995foundations}. In recent years, its popularity has expanded beyond database applications to a diverse range of data-intensive domains, including program analysis~\cite{scholz2016fast}, reverse engineering~\cite{flores2020datalog}, graph analytics~\cite{shkapsky2016big}, and machine learning~\cite{li2023scallop}. A Datalog program consists of a collection of logical rules, each written as a Horn clause:
\[
\begin{array}{lcl}
 \textit{Head}(...)  & \leftarrow & \textit{Body}_1(...), ..., \textit{Body}_n(...).
\end{array}
\]
The head (left-hand side of $\leftarrow$) is a single predicate atom representing a derived fact, and the body (right-hand side of $\leftarrow$) is a conjunction of predicate atoms specifying the conditions under which the head holds. Shared logical variables across body atoms impose equality constraints, directly corresponding to equi-join conditions in relational algebra. A Datalog program partitions its data into an extensional database (EDB) of base facts, provided as input, and an intensional database (IDB) of derived facts, produced by rule application. Evaluation proceeds by iteratively applying all rules until a fixpoint is reached, that is, until no new IDB facts can be derived~\cite{ceri1989you}. This iterative fixpoint semantics distinguishes Datalog from traditional SQL: while both share relational foundations, Datalog is designed to support recursion seamlessly, enabling transitive inference.
For example, the following transitive closure query derives all pairs of
nodes connected by a directed path:
\[
\begin{array}{lcl}
 \textit{TC}(X,Y) & \leftarrow & \textit{Edge}(X,Y). \\
 \textit{TC}(X,Y) & \leftarrow & \textit{TC}(X,Z),\; \textit{Edge}(Z,Y).
\end{array}
\]
Each rule is compiled into relational
algebra~\cite{ceri1986translation}, and recursion is handled by a
least-fixpoint operator (denoted $\mathcal{O}$) that iteratively applies the resulting expressions.
The transitive closure compiles to:
\[
\textit{TC} \;=\; \mathcal{O}_{\textit{TC}}\Bigl(\,
  \textit{Edge}
  \;\cup\;
  \pi_{X,Y}\bigl(\textit{TC} \bowtie_{Z} \textit{Edge}\bigr)
\,\Bigr)
\]
To avoid redundant derivation, modern engines use \emph{semi-na\"ive
evaluation}~\cite{balbin1987generalization}, which maintains a delta
relation $\Delta\textit{TC}^{(i)}$ of the tuples derived in the previous iteration and restricts the next iteration to joining only this delta relation against existing facts:
\[
\Delta\textit{TC}^{(i+1)} \;=\;
  \pi_{X,Y}\bigl(\Delta\textit{TC}^{(i)} \bowtie_{Z} \textit{Edge}\bigr)
  \;\setminus\; \textit{TC}^{(i)}
\]
This incremental strategy is the de facto standard in Datalog
engines and forms the basis of our evaluation approach.

\subsection{Worst-Case Optimal Joins (WCOJ)}
Efficient join processing is central to the performance of a Datalog engine.
While acyclic queries can be evaluated efficiently using semijoin reductions~\cite{yannakakis1981algorithms}, the cyclic multi-way queries common in graph analytics and recursive Datalog expose a fundamental weakness of execution based on binary join plans.
The triangle query illustrates this issue:
$\textit{Triangle}(x,y,z) \leftarrow R(x,y),\ S(y,z),\ T(z,x)$. A traditional binary join plan processes this query by scanning the tuples of the leftmost relation ($R$) and probing indexes on the remaining relations, as shown in the pseudocode of Figure~\ref{fig:bj-vs-wcoj-code}(a). Figure~\ref{fig:binary-vs-wcoj} illustrates why this tuple-at-a-time execution is costly. Consider a scenario in which relations $R$ and $S$ form a dense star around a common vertex ($y = 1$). For the first tuple of $R$ ($x = 1$), the nested loops blindly enumerate all open two-hop paths through $y = 1$, materializing 4 intermediate rows. When this is repeated for the remaining tuples of $R$ ($x=2,3,4$), the intermediate table grows to 16 rows in total. For large, dense graphs, the size of this binary intermediate can reach $O(|E|^2)$. Yet when this intermediate table is finally probed against $T$ (the \texttt{if} statement in Figure~\ref{fig:bj-vs-wcoj-code}(a)), 15 of the 16 paths are discarded because the closing edge does not exist.

  \begin{figure}[t]
  \centering\small
  \begin{lstlisting}[
    mathescape=true,
    basicstyle=\small\ttfamily,
    xleftmargin=1em,
    numbers=none
  ]
  // (a) Binary join plan [R, S, T]
  for (x, y) in R:          $\triangleright$ scan
    for z in S[y]:           $\triangleright$ probe
      temp.emit(x, y, z)
  for (x, y, z) in temp:
    if (z, x) in T:        $\triangleright$ probe
      Triangle.emit(x, y, z)

  // (b) WCOJ plan [x, y, z]
  for x in R.x $\cap$ T.x:          $\triangleright$ intersect
    for y in R[x].y $\cap$ S.y:     $\triangleright$ intersect
      for z in S[y].z $\cap$ T[x].z: $\triangleright$ intersect
        Triangle.emit(x, y, z)
  \end{lstlisting}
  \caption{Chained binary join vs.\ WCOJ for the triangle query.
    }
  \label{fig:bj-vs-wcoj-code}
  \vspace{-0.3cm}
  \end{figure}

This exposes the fundamental problem of binary joins on cyclic workloads: they generate intermediate results that far exceed the size of the final output. The AGM bound~\cite{atserias2013size} shows that the output of the triangle query contains at most $O(|E|^{3/2})$ tuples, whereas the intermediate results of a binary join plan can reach $O(|E|^{2})$. Worst-Case Optimal Joins (WCOJ) are a family of algorithms designed specifically to run within the AGM bound. WCOJ abandons tuple-at-a-time processing in favor of an attribute-at-a-time approach. As shown in the pseudocode of Figure~\ref{fig:bj-vs-wcoj-code}(b), the algorithm cross-references relations dynamically, intersecting $R.x$ and $T.x$ to isolate productive $x$ bindings before evaluating $y$ and $z$. As illustrated in Figure~\ref{fig:binary-vs-wcoj}, this aggressive early pruning discards the 15 nonviable branches before they are ever materialized, ensuring that only the viable output tuple $(1,1,1)$ is processed.

Following the formalization of the WCOJ framework~\cite{ngo2018worst}, several single-node implementations emerged. Leapfrog Triejoin (LFTJ)~\cite{veldhuizen2014leapfrog} operates over sorted tries using linear-search-based iteration, while Free Join~\cite{wang2023free} unifies binary and WCOJ plans via lazy trie materialization. HoneyComb~\cite{wu2025honeycomb} further advanced the state of the art by introducing a partition-based parallel WCOJ execution model; however, like its predecessors, HoneyComb targets single-node, shared-memory architectures.
Distributed WCOJ has so far been studied for one-shot conjunctive and subgraph queries on dataflow and big-data platforms~\cite{chu2015theory,ammar2018distributed,zhang2021fast} (Section~\ref{sec:related}); scaling WCOJ-based \emph{recursive} evaluation to shared-nothing HPC clusters remains an open challenge. The attribute-at-a-time iteration required by WCOJ is highly sensitive to network latency. If data is naively distributed across a cluster, these prefix-extension loops trigger millions of synchronous remote probes, replacing the intermediate-shuffle bottleneck with a crippling network-latency bottleneck. Sections~\ref{sec:par-wcoj} and~\ref{iter_wcoj_par} describe how our design avoids these problems.

\begin{figure*}[t]
    \centering
    \includegraphics[width=\textwidth]{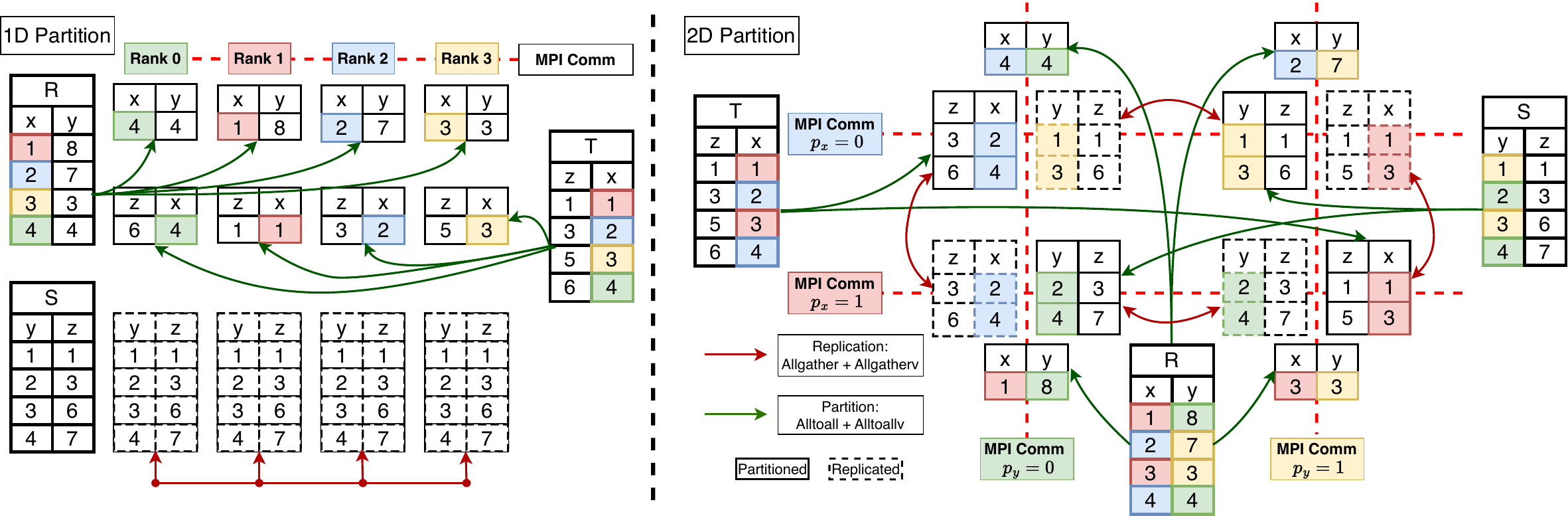}
    \caption{1D versus 2D partitioning for the triangle query $\textit{Triangle}(x,y,z) \leftarrow R(x,y), S(y,z), T(z,x)$. The 1D layout partitions tuples on a single key, which forces replication of relations that lack that key. The 2D layout partitions each relation on the keys it contains and replicates it along the missing dimensions. Green arrows indicate partitioning, and red arrows indicate replication.}
    \vspace{-0.3cm}
    \label{fig:1d_2d_partition}
\end{figure*}

\subsection{Distributed Datalog Processing}
Several systems have extended Datalog evaluation to distributed clusters, including BigDatalog~\cite{shkapsky2016big} and RaSQL~\cite{gu2019rasql}, both built atop Apache Spark, and Distributed SociaLite~\cite{seo2013socialite}, which uses its own master--worker runtime. While these engines can process datasets that exceed single-machine memory capacity, their scaling efficiency remains limited, often exhibiting order-of-magnitude slowdowns relative to optimized single-node engines such as Soufflé~\cite{scholz2016fast}. Slog~\cite{gilray2024datalog} represents a significant step forward as the first Datalog engine designed explicitly for shared-nothing multi-node execution, built on an MPI-based parallel relational algebra framework~\cite{gilray2021compiling}. Nevertheless, like prior implementations, Slog relies on materialized binary joins, leaving it susceptible to intermediate result explosion under unfavorable join orders. On queries with complex or cyclic join patterns, its scalability degrades substantially despite efficient communication and load balancing. Consequently, no existing distributed Datalog engine is robust across the full range of queries encountered in practice.

\section{Workload Partitioning for Distributed Worst-Case Optimal Joins}
\label{sec:par-wcoj}

Parallel multi-way join execution in both shared-memory and shared-nothing systems has traditionally relied on chained binary joins, which materialize intermediate relations that can grow prohibitively large for complex, cyclic, or recursive queries. In shared-nothing environments, this intermediate blowup is particularly costly, as the system must repeatedly redistribute these oversized relations across processes, inflating both computation and communication overhead.
While WCOJ is well established as a solution to the intermediate blowup inherent in chained binary joins (Figure~\ref{fig:binary-vs-wcoj}, right), its distributed execution has so far been studied only for one-shot, non-recursive queries~\cite{chu2015theory,ammar2018distributed}. The central challenge is to identify a partitioning strategy that minimizes inter-process communication while supporting the continuous set intersection across all join variables that WCOJ requires, and that can be reapplied cheaply to the newly derived tuples of every fixpoint iteration. We build on the HyperCube (Shares) algorithm~\cite{afrati2010optimizing,beame2017communication}, in which join variables define the dimensions of a logical process grid, and relations are hash-partitioned along the dimensions corresponding to their attributes and selectively replicated along the dimensions whose attributes they do not carry, and we realize it with MPI collectives over subcube communicators. This design maximizes the data locality that is essential for efficient parallel join execution while minimizing communication overhead. The remainder of this section formalizes the partitioning scheme, describes its MPI-based implementation, derives a cost model characterizing the resulting replication overhead, and analyzes the impact of partition dimensionality.

\subsection{Hypercube Partitioning and Replication for Distributed WCOJ}
\label{sec:hypercube_partition}

The hypercube partitioning scheme distributes relations across all processes to support distributed WCOJ. To motivate the approach, we first review the traditional radix-hash partitioning used in chained binary join execution. In that setting, every join involves exactly two relations, and the standard strategy is to partition each relation by applying a radix-hash function to the join column(s). This guarantees that all tuples sharing a join key are co-located on the same process, enabling fully local, communication-free parallel join execution. For instance, given $R(x,y) \bowtie S(y,z)$, both $R$ and $S$ are partitioned on column $y$ using the same radix-hash function, ensuring that matching tuples reside on the same process and that the join proceeds without data movement. This strategy is inherently relation-centric: partitioning decisions are made independently for each pair of relations involved in a binary join.

WCOJ, by contrast, is variable-centric: when joining more than two relations, it requires simultaneous set intersection across all join variables. Consider the triangle query $R(x,y) \bowtie S(y,z) \bowtie T(z,x)$: relation $R$ must be intersected with both $S$ (on column $y$) and $T$ (on column $x$), which makes the traditional radix-hash strategy inapplicable. Since partitioning decisions must now be made with respect to variables rather than individual relation pairs, an additional degree of freedom emerges: one may partition on one, two, or all three variables, each choice yielding a distinct trade-off. A natural way to implement this variable-centric partitioning is a hypercube layout~\cite{afrati2010optimizing,beame2017communication}, in which selected join variables define the grid dimensions and tuples are partitioned or replicated according to the variables they contain.

\paragraph{Notation and Setup}
Consider a conjunctive query over relations \(R_1, \dots, R_n\).
The variables shared across multiple relations dictate the query
topology and act as \emph{join keys}. To distribute
evaluation, we select a subset of these join keys as orthogonal
\emph{partition keys}, denoted by \(D = \{d_1, \dots, d_k\}\).
These partition keys define a \(k\)-dimensional logical
hypercube. If dimension \(d_i \in D\) is divided into \(p_i\)
partitions, the resulting process grid contains \(P = \prod_{i=1}^{k} p_i\)
cells, with each cell mapped to one MPI rank. In the runtime, all ranks belong to a global MPI communicator that spans the full process grid. In addition, the runtime
creates subcube communicators for individual dimensions, which
are used to replicate tuples only along the missing dimensions. For example, in the 2D layout of Figure~\ref{fig:1d_2d_partition}, four ranks are organized as a 2$\times$2 grid, and each row and each column has its own MPI subcommunicator.

\paragraph{Algorithm and Implementation}
Hypercube partitioning organizes processes as a logical grid whose
dimensions are defined by the selected partition keys. For each
tuple \(t\) in a relation \(R\), the runtime first creates a
coordinate vector \(\textit{Coord}=[\star,\dots,\star]\). For
every selected partition key
\(d_j \in D \cap \mathrm{keys}(R)\), it computes
\(\textit{Coord}[j] = t[d_j] \bmod p_j\).
Figure~\ref{fig:1d_2d_partition} illustrates this routing in a 2D logical grid partitioned on variables $x$ and $y$, where each dimension is divided into two partitions. Given the tuple $(1, 8)$ of $R(x,y)$, the system computes its grid coordinates as $(1 \bmod 2, \; 8 \bmod 2) = (1, 0)$, i.e., the rank in row $p_x = 1$ and column $p_y = 0$.

Using the computed \textit{Coord}, the runtime routes each tuple
according to whether the relation fully covers the selected
partition-key set \(D\). If \(D \subseteq \mathrm{keys}(R)\), all
entries of \textit{Coord} are determined, so the tuple maps to
exactly one logical cell, which corresponds to a unique rank
through \textsc{CoordToRank}. In this case, the relation is
fully partitioned and no replication is required. In the MPI
implementation, this redistribution is realized over the global
communicator by first exchanging per-destination tuple counts
with \texttt{MPI\_Alltoall}, followed by the actual tuple
payload exchange with \texttt{MPI\_Alltoallv}.

Otherwise, some selected partition keys are missing from
\(R\), so some entries of \textit{Coord} remain unspecified.
In this case, the known entries determine a partial coordinate
instead of a unique rank. The tuple is first grouped with all
tuples that share the same known coordinate, and then
replicated across all ranks whose coordinates are consistent
with that partial assignment. Thus, partitioning is applied on
the known dimensions, and replication is applied only along the
missing dimensions. This is exactly the behavior captured by
\textit{MissingDims}, \textit{TgtCoords}, and \textit{TgtRanks}
in Procedure~\ref{alg:relation_partition_replication}. In the
MPI implementation, the runtime still begins with a global
\texttt{MPI\_Alltoall}+\texttt{MPI\_Alltoallv} redistribution
based on the known dimensions, so that tuples with the same
partial coordinate are co-located in the same
subcube. It then performs selective replication inside the
relevant row, column, or higher-dimensional subcube
communicator by first exchanging counts with
\texttt{MPI\_Allgather} and then replicating payloads with
\texttt{MPI\_Allgatherv}. When multiple dimensions are missing,
this replication is applied one missing dimension at a time.

Figure~\ref{fig:1d_2d_partition} shows this placement process in
detail under both 1D and 2D layouts. In the 1D case, a tuple is
either partitioned along the selected key, if that key is
present in the relation, or replicated across all ranks if the
relation does not contain that key. In the 2D case, a relation
containing both selected partition keys is fully partitioned, so
each tuple is sent to one unique target cell. By contrast,
relations containing only one of the two selected keys are first
partitioned along the known dimension and then replicated along
the missing one. For example, as shown in
Figure~\ref{fig:1d_2d_partition}, with partition keys \(x\) and
\(y\), tuple \((1,1)\) in \(T(z,x)\) is assigned using its
\(x\)-value and then replicated across all ranks that share that
\(x\)-coordinate, since \(y\) is absent from the relation. In
the figure, this appears as an initial placement into row- or
column-aligned groups, followed by replication across the
corresponding communicator. In the implementation, the initial
placement corresponds to the global redistribution phase, while
the row- or column-wise expansion is realized by
\texttt{MPI\_Allgather}+\texttt{MPI\_Allgatherv} within the
corresponding subcube communicator. This design avoids fine-grained
remote lookups during WCOJ execution while still preserving complete
joinability locally.

\begin{algorithm}[t]
\small
\caption{Relation Partition and Replication for Local WCOJ}
\label{alg:relation_partition_replication}
\begin{algorithmic}[1]
\Require $Q=\{R_1,\dots,R_m\}$
\Require $D=\{d_1,\dots,d_k\}$
\Require $\{p_1,\dots,p_k\}$
\Ensure Target-rank assignment for local WCOJ

\ForAll{$R_i \in Q$}
    \ForAll{$t \in R_i$}
        \State // Construct coordinate based on partition keys $D$
        \State $\emph{Coord} \gets [\star,\dots,\star]$
        \State $\emph{MissingDims} \gets D \setminus \emph{keys}(R_i)$
        \State // Fill in coordinates from the partition-key
        \State // values present in the tuple
        \ForAll{$d_j \in D \cap \emph{keys}(R_i)$}
            \State $\emph{Coord}[j] \gets t[d_j] \bmod p_j$
        \EndFor

        \State // If there is no missing dimension
        \If{$\emph{MissingDims} = \emptyset$}
            \State $r \gets \textsc{CoordToRank}(Coord)$
            \State Send $(R_i,t)$ to rank $r$
        \Else
            \State // Enumerate coordinates matching the partial one
            \State $\emph{TgtCoords} \gets \{(c_1,\dots,c_k)\mid$
            \State \hspace{1.2em} $c_j = Coord[j]$ if $Coord[j]\neq \star,$
            \State \hspace{1.2em} $0 \le c_j < p_j$ otherwise $\}$
            \State // All ranks consistent with the partial coordinate
            \State $\emph{TgtRanks} \gets \{\textsc{CoordToRank}(c)\mid$
            \State \hspace{1.2em} $c\in \emph{TgtCoords}\}$
            \State Replicate $(R_i,t)$ to all ranks in $\emph{TgtRanks}$
        \EndIf
    \EndFor
\EndFor
\State Materialize local fragments
\State Build local indexes
\State Run local WCOJ
\end{algorithmic}
\end{algorithm}

\subsection{Cost Model for Replication under Hypercube Partitioning}
\label{sec:cost_model_replication}

We now characterize the replication cost induced by hypercube partitioning, restating the standard Shares analysis~\cite{afrati2010optimizing} in our notation.
Consider a query \(Q=\{R_1,R_2,\dots,R_n\}\) with join-key set \(J\).
The system selects partition keys
\(D=\{d_1,d_2,\dots,d_k\}\subseteq J\), where dimension \(d_j\)
has \(p_j\) partitions. The resulting \(k\)-dimensional hypercube contains $P=\prod_{j=1}^{k} p_j$ logical cells.
For each relation \(R_i\in Q\), let \(\mathrm{keys}(R_i)\) denote the join keys
appearing in \(R_i\). If \(D\subseteq \mathrm{keys}(R_i)\), then \(R_i\)
is fully partitioned: each tuple maps to exactly one hypercube cell and no
replication is needed. Otherwise, \(R_i\) is only partially covered and must
be replicated along the missing dimensions
\(D\setminus \mathrm{keys}(R_i)\).

We define the replication factor of \(R_i\) as
\[
\gamma_i
= \prod_{d_j \in D \setminus \mathrm{keys}(R_i)} p_j,
\]
which is the number of hypercube cells containing each original tuple of
\(R_i\). If the product is empty, then \(\gamma_i=1\).

Let \(S_i\) denote the original size of \(R_i\), measured either in tuples
or in bytes. After partitioning and replication, the total and
extra replicated sizes of \(R_i\) are
\[
S_i^{\mathrm{rep}} = S_i \gamma_i, \qquad
S_i^{\mathrm{extra}} = S_i(\gamma_i - 1).
\]

Summing over all relations gives the total replicated size of the query:
\begin{align}
S_Q^{\mathrm{rep}}
  &= \sum_{i=1}^{n} S_i^{\mathrm{rep}} = \sum_{i=1}^{n} S_i \gamma_i.
\end{align}

This model captures the main trade-off of hypercube partitioning:
relations covering more partition keys incur less replication, whereas
relations missing more partition dimensions incur larger replication factors,
especially when the corresponding \(p_j\) are large. Hence, the choice of
partition keys determines both per-relation replication and the total
replication cost of the query.

\subsection{Impact of Partition Dimensionality}
\label{1d2d3d}
Partition dimensionality is a central design choice in distributed
WCOJ because it determines both tuple co-location for local
prefix-based evaluation and the replication needed to preserve
complete joinability. As shown by the cost model above, replication
is governed not only by the number of partition dimensions, but more
fundamentally by the alignment between the selected partition keys
and the attributes present in each relation. Increasing the number
of partition dimensions can improve locality for local join
evaluation, but it can also increase communication complexity and
replication along dimensions that a relation does not cover.

In general, lower-dimensional partitioning provides a simpler data
layout and communication pattern, but does not necessarily reduce
replication. When the selected partition keys align well with a
relation schema, replication can be small or avoided entirely.
However, if a relation misses one of the chosen partition keys, it
must still be replicated along that dimension, and under some
layouts this effectively amounts to broad replication. By contrast,
higher-dimensional partitioning often allows each relation to match
at least part of the selected key set, which can improve data
placement and avoid wholesale replication. The trade-off
is that more relations become only partially covered and
must be replicated along their missing dimensions, leading to a more
complex communication pattern. The net effect is inherently
query-dependent, determined by how the chosen partition
dimensions match the schemas of the participating relations.

We illustrate this trade-off using the \textsc{Peer Influence} query:
\[
\begin{array}{lcl}
\texttt{Influenced}(x) & \leftarrow & \texttt{Seed}(x). \\
\texttt{Influenced}(y) & \leftarrow & \texttt{Influenced}(x),\, R(x,y),\, \\
 & & S(x,z),\, T(z,y).
\end{array}
\]
where \(R\), \(S\), and \(T\) denote the three logical occurrences of
the same base relation \texttt{Edge}. This query involves three
variables, \(x\), \(y\), and \(z\). For the cost analysis, let
\(N_I = |\texttt{Influenced}|\) denote the size of relation
\texttt{Influenced}, and let
\[
N_R = N_S = N_T = N_E
\]
denote the size of each logical edge atom, since all three refer to
the same underlying base relation \texttt{Edge}. We use \(R\), \(S\),
and \(T\) only to distinguish the attribute coverage of the three
logical atoms in the join, while the physical relation itself remains
the same \texttt{Edge} relation. We next compare 1D, 2D, and 3D
partitioning under the same total number of ranks.

\paragraph{1D partitioning}
We first consider a 1D layout with \(64\) ranks and partition key
\(D=\{x\}\). Under this choice, \(\texttt{Influenced}(x)\),
\(R(x,y)\), and \(S(x,z)\) all contain the selected key and are
therefore fully partitioned. In contrast, \(T(z,y)\) does not contain
\(x\), so it cannot be partitioned on the chosen dimension and must be
replicated across all \(64\) partitions. As shown in
Table~\ref{tab:peer-dim-examples}, this layout is simple, but it
imposes a high replication burden on the uncovered relation.

\paragraph{2D partitioning}
Next, we consider a 2D layout with an \(8\!\times\!8\) grid and \(D\!=\!\{x,z\}\). Here, only \(S(x,z)\) contains both
selected dimensions and is therefore fully partitioned.
\(\texttt{Influenced}(x)\) and \(R(x,y)\) lack \(z\), while \(T(z,y)\)
lacks \(x\), so these relations are partitioned on the dimension they
contain and replicated along the missing one. Compared with 1D,
replication is no longer concentrated in a single relation but is
distributed more evenly across the partially covered relations.

\paragraph{3D partitioning}
Finally, we consider a 3D layout with a \(4\times4\times4\) grid and
partition keys \(D=\{x,y,z\}\). In this layout,
\(\texttt{Influenced}(x)\) contains only one selected dimension and
must be replicated along the other two, whereas each binary edge atom
contains two selected dimensions and is replicated along only one.
Thus, 3D partitioning gives the finest-grained placement for the edge
relations, but shifts a larger share of the replication burden onto
the unary relation \(\texttt{Influenced}\).

\begin{table*}[t]
\centering
\scriptsize
\caption{Effect of partition dimensionality on \textsc{Peer Influence} with 64 total ranks. Blue indicates partitioned dimensions; red indicates replication along missing selected dimensions. A2A+A2AV denotes \texttt{MPI\_Alltoall}+\texttt{MPI\_Alltoallv}, and AG+AGV denotes \texttt{MPI\_Allgather}+\texttt{MPI\_Allgatherv}.}
\label{tab:peer-dim-examples}
\setlength{\tabcolsep}{2.4pt}

\begin{minipage}[t]{0.31\textwidth}
\centering
\textbf{1D: \(D=\{x\}\), \(64\) parts.}\par\vspace{2pt}
\begin{tabular}{lcccc}
\toprule
\textbf{Rel.} & \(x(64)\) & \textbf{Sto.} & \makecell{\#A2A\\+A2AV} & \makecell{\#AG\\+AGV} \\
\midrule
\(\texttt{Inf.}(\textcolor{blue}{x})\)
& \textcolor{blue}{P} & \(N_I\) & 1 & 0 \\

\(R(\textcolor{blue}{x},y)\)
& \textcolor{blue}{P} & \(N_E\) & 1 & 0 \\

\(S(\textcolor{blue}{x},z)\)
& \textcolor{blue}{P} & \(N_E\) & 1 & 0 \\

\(T(z,y)\)
& \textcolor{red}{\(\times 64\)} & \(64N_E\) & 0 & 1 \\
\bottomrule
\end{tabular}
\end{minipage}
\hfill
\begin{minipage}[t]{0.33\textwidth}
\centering
\textbf{2D: \(D=\{x,z\}\), \(8\times8\) grid.}\par\vspace{2pt}
\begin{tabular}{lccccc}
\toprule
\textbf{Rel.} & \(x(8)\) & \(z(8)\) & \textbf{Sto.} & \makecell{\#A2A\\+A2AV} & \makecell{\#AG\\+AGV} \\
\midrule
\(\texttt{Inf.}(\textcolor{blue}{x})\)
& \textcolor{blue}{P} & \textcolor{red}{\(\times 8\)} & \(8N_I\) & 1 & 1 \\

\(R(\textcolor{blue}{x},y)\)
& \textcolor{blue}{P} & \textcolor{red}{\(\times 8\)} & \(8N_E\) & 1 & 1 \\

\(S(\textcolor{blue}{x},\textcolor{blue}{z})\)
& \textcolor{blue}{P} & \textcolor{blue}{P} & \(N_E\) & 1 & 0 \\

\(T(\textcolor{blue}{z},y)\)
& \textcolor{red}{\(\times 8\)} & \textcolor{blue}{P} & \(8N_E\) & 1 & 1 \\
\bottomrule
\end{tabular}
\end{minipage}
\hfill
\begin{minipage}[t]{0.34\textwidth}
\centering
\textbf{3D: \(D=\{x,y,z\}\), \(4\times4\times4\) grid.}\par\vspace{2pt}
\begin{tabular}{lcccccc}
\toprule
\textbf{Rel.} & \(x(4)\) & \(y(4)\) & \(z(4)\) & \textbf{Sto.} & \makecell{\#A2A\\+A2AV} & \makecell{\#AG\\+AGV} \\
\midrule
\(\texttt{Inf.}(\textcolor{blue}{x})\)
& \textcolor{blue}{P} & \textcolor{red}{\(\times 4\)} & \textcolor{red}{\(\times 4\)} & \(16N_I\) & 1 & 2 \\

\(R(\textcolor{blue}{x},\textcolor{blue}{y})\)
& \textcolor{blue}{P} & \textcolor{blue}{P} & \textcolor{red}{\(\times 4\)} & \(4N_E\) & 1 & 1 \\

\(S(\textcolor{blue}{x},\textcolor{blue}{z})\)
& \textcolor{blue}{P} & \textcolor{red}{\(\times 4\)} & \textcolor{blue}{P} & \(4N_E\) & 1 & 1 \\

\(T(\textcolor{blue}{z},\textcolor{blue}{y})\)
& \textcolor{red}{\(\times 4\)} & \textcolor{blue}{P} & \textcolor{blue}{P} & \(4N_E\) & 1 & 1 \\
\bottomrule
\end{tabular}
\end{minipage}

\end{table*}
Overall, these three cases illustrate that the effect of partition
dimensionality is governed by how the selected partition keys align
with the schemas of the participating relations. In this query, 1D
partitioning leads to very large replication for the relation that
does not contain the chosen key, 2D partitioning spreads the
replication more evenly across partially covered relations, and 3D
partitioning further reduces the replication of each binary atom at
the cost of increasing replication for the unary relation. The best
layout is therefore not determined by dimensionality alone, but by
the interaction between query structure, relation schemas, and
partition choice.

\begin{figure}[t]
    \centering
    \includegraphics[width=\linewidth]{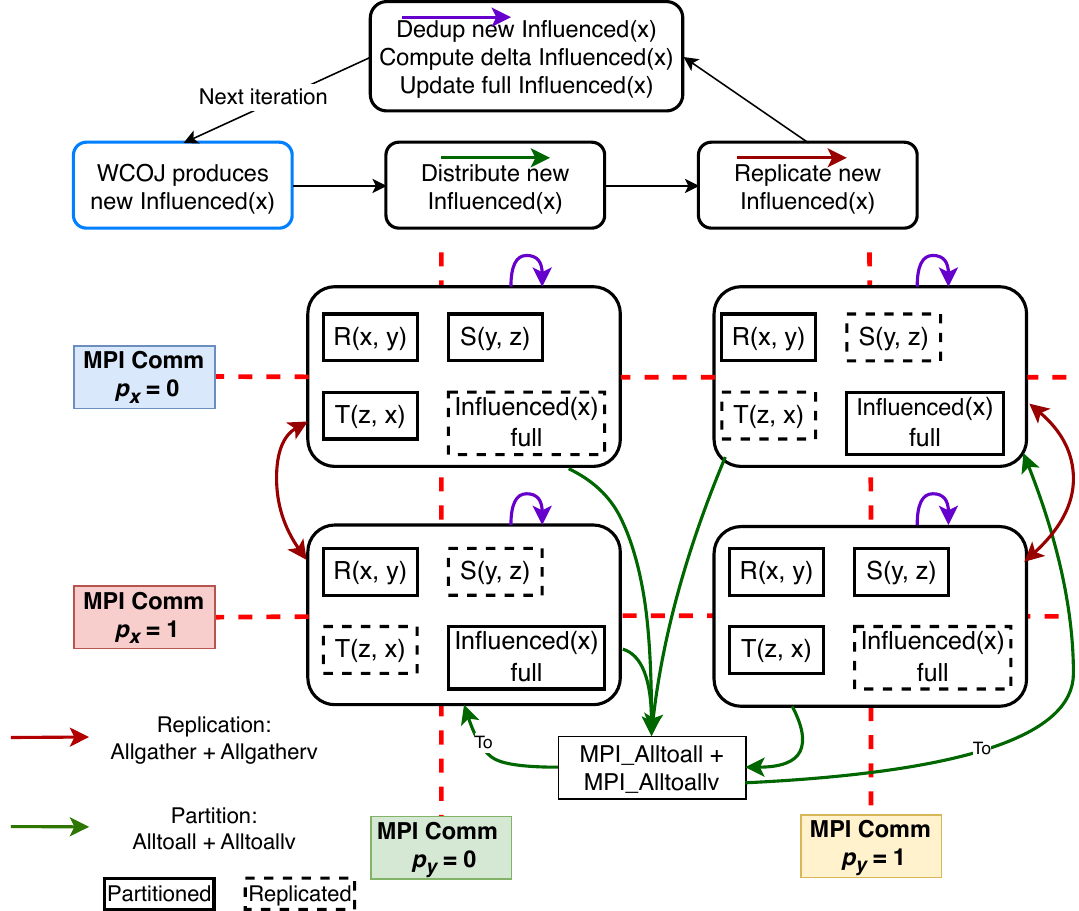}
    \caption{Iterative distributed WCOJ for \textsc{Peer Influence} under a 2D hypercube partition. Newly derived facts are redistributed across ranks by \texttt{MPI\_Alltoall(v)}, replicated along missing dimensions by \texttt{MPI\_Allgather(v)} when needed, and then deduplicated to form the delta for the next iteration. Solid boxes denote partitioned data, and dashed boxes denote replicated data.}
    \label{fig:iter-wcoj-flow}
    \vspace{-0.4cm}
\end{figure}

\section{Iterative WCOJ in Parallel}
\label{iter_wcoj_par}

While non-recursive queries such as the Triangle query require only a single pass of multi-way join evaluation, recursive queries such as \textit{Peer Influence} require repeated evaluation until a global fixpoint is reached. In the \textit{Peer Influence} query, the head relation \texttt{Influenced} appears in the body, meaning each iteration derives new influenced vertices that may in turn trigger further derivations in subsequent iterations. To handle this efficiently, our system adopts semi-na\"ive fixpoint evaluation, in which each iteration derives new facts from the current delta relation $\Delta^{(i)}$ rather than re-evaluating the entire recursive state from scratch. Each iteration is driven by a full round of distributed WCOJ. Prior HyperCube-based systems~\cite{chu2015theory,zhang2021fast} perform a single shuffle before a one-shot join; recursive evaluation instead requires the partitioning to be reapplied to the newly derived tuples of every iteration. Section~\ref{sec:par-wcoj} described how multidimensional hypercube partitioning and selective replication organize relation fragments so that WCOJ can proceed locally at each rank in a single round. This section explains how that distributed WCOJ execution is extended to the iterative setting, such that the output of iteration $i$ becomes the input delta $\Delta^{(i+1)}$ for the next iteration, continuing until the global delta is empty and a fixpoint is reached.

\paragraph{Overall Iterative WCOJ Framework}
At iteration $i$, each rank evaluates the recursive rules using the current delta relation $\Delta^{(i)}$ together with its local full relation state. WCOJ is executed locally over these structures to derive candidate new tuples without materializing any large intermediate relations. After local evaluation, newly derived tuples are routed to their owner ranks according to the hypercube partitioning scheme, checked against existing state, and retained only if they are genuinely new. The surviving tuples form $\Delta^{(i+1)}$, which is merged into the full relation state before the next round begins. Iteration terminates when a global reduction confirms that the delta is empty across all ranks, at which point the fixpoint has been reached.

\paragraph{Maintaining Local WCOJ Execution Across Iterations}
The key invariant that our design preserves across iterations is that the inner prefix-driven intersection work of WCOJ remains entirely rank-local. If relation fragments required for prefix extension were allowed to reside on remote ranks, WCOJ would trigger fine-grained point-to-point remote lookups during the intersection loop, introducing synchronous communication latency that would negate the computational benefits of WCOJ. Our design avoids this by reusing the same partition-and-replication layout established in Section~\ref{sec:par-wcoj} for every iteration. All communication is deferred to the redistribution phase that follows local evaluation. As shown in Figure~\ref{fig:iter-wcoj-flow}, after each rank completes its local WCOJ pass and produces a set of candidate new tuples, these tuples are redistributed using the same hypercube partitioning scheme applied to the input relations. Specifically, the runtime first uses \texttt{MPI\_Alltoall} and \texttt{MPI\_Alltoallv} to route tuples to their owner ranks based on the partitioned dimensions they carry. Where replication is required along missing dimensions, the runtime subsequently issues \texttt{MPI\_Allgather} and \texttt{MPI\_Allgatherv} calls within the appropriate subcube communicators to propagate tuples along those dimensions. The resulting new delta is then deduplicated against the existing local state at each responsible rank, retaining only genuinely new tuples for the next iteration.
This design cleanly separates local computation from inter-rank communication, preserving the locality and efficiency of rank-local multi-way join execution across all iterations while guaranteeing correct distributed fixpoint progress.

\begin{table*}[t]
\centering
\small
\setlength{\tabcolsep}{4pt}
\caption{Runtime (seconds) on Polaris under shared-nothing execution for 2D and 3D partitioning. Lower is better.}
\label{tab:polaris-2d-3d}
\begin{tabular}{lcccccccccccc}
\toprule
& \multicolumn{2}{c}{\makecell{\textbf{Peer}\\\textbf{(LiveJournal)}\\\textbf{undirected}}}
& \multicolumn{2}{c}{\makecell{\textbf{Peer}\\\textbf{(Orkut)}\\\textbf{undirected}}}
& \multicolumn{2}{c}{\makecell{\textbf{Triangle}\\\textbf{(LiveJournal)}\\\textbf{undirected}}}
& \multicolumn{2}{c}{\makecell{\textbf{Triangle}\\\textbf{(Orkut)}\\\textbf{undirected}}}
& \makecell{\textbf{Andersen}\\\textbf{(Large)}}
& \makecell{\textbf{Andersen}\\\textbf{(Medium)}}
& \makecell{\textbf{TC}\\\textbf{(USRoad)}}
& \makecell{\textbf{SG}\\\textbf{(G10)}} \\
{\scriptsize\textbf{Input edges}} & \multicolumn{2}{c}{{\scriptsize 86,220,856 + 10 seeds}} & \multicolumn{2}{c}{{\scriptsize 234,370,166 + 10 seeds}} & \multicolumn{2}{c}{{\scriptsize 86,220,856}} & \multicolumn{2}{c}{{\scriptsize 234,370,166}} & {\scriptsize 67,999,999} & {\scriptsize 33,999,999} & {\scriptsize 165,435} & {\scriptsize 99,805} \\
{\scriptsize\textbf{Output edges}} & \multicolumn{2}{c}{{\scriptsize 3,727,171}} & \multicolumn{2}{c}{{\scriptsize 2,936,543}} & \multicolumn{2}{c}{{\scriptsize 1,793,261,106}} & \multicolumn{2}{c}{{\scriptsize 3,765,505,086}} & {\scriptsize 5,315,019} & {\scriptsize 2,654,871} & {\scriptsize 871,365,688} & {\scriptsize 100,000,000} \\
\cmidrule(lr){2-3}
\cmidrule(lr){4-5}
\cmidrule(lr){6-7}
\cmidrule(lr){8-9}
\textbf{Nproc} & \textbf{3D} & \textbf{2D}
& \textbf{3D} & \textbf{2D}
& \textbf{3D} & \textbf{2D}
& \textbf{3D} & \textbf{2D}
& \textbf{2D}
& \textbf{2D}
& \textbf{2D}
& \textbf{2D} \\
\midrule
32   & 23.896 & 15.576 & 68.247 & OOM    & 28.516 & 27.036 & 91.546 & OOM    & 17.519 & 7.191 & 20.199 & 78.235 \\
64   & 23.259 & 11.253 & 68.945 & 30.765 & 18.786 & 15.786 & 61.128 & 48.958 & 8.169  & 4.164 & 10.336 & 41.947 \\
128  & 13.712 & 6.796  & 39.923 & 19.378 & 12.992 & 10.042 & 42.758 & 30.195 & 6.794  & 3.315 & 6.042  & 23.023 \\
256  & 10.972 & 5.530  & 29.822 & 16.366 & 9.314  & 7.037  & 29.914 & 22.638 & 4.198  & 2.002 & 3.596  & 13.169 \\
512  & 11.729 & 3.688  & 39.556 & 10.087 & 7.824  & 5.153  & 29.515 & 15.886 & 3.804  & 1.852 & 2.495  & 7.834  \\
1024 & 6.812  & 4.173  & 26.301 & 10.131 & 7.444  & 5.128  & 30.246 & 17.053 & 2.676  & 1.552 & 3.515  & 6.975  \\
2048 & 6.366  & 3.243  & 23.317 & 7.955  & 5.614  & 4.363  & 29.971 & 12.207 & 3.141  & 1.786 & 7.336  & 5.462  \\
\bottomrule
\end{tabular}
\end{table*}

\begin{table*}[t]
\centering
\small
\setlength{\tabcolsep}{4pt}
\caption{Runtime (seconds) on Aurora under shared-nothing execution for 2D and 3D partitioning. Lower is better.}
\label{tab:aurora-2d-3d}
\begin{tabular}{lcccccccccccc}
\toprule
& \multicolumn{2}{c}{\makecell{\textbf{Peer}\\\textbf{(LiveJournal)}\\\textbf{undirected}}}
& \multicolumn{2}{c}{\makecell{\textbf{Peer}\\\textbf{(Orkut)}\\\textbf{undirected}}}
& \multicolumn{2}{c}{\makecell{\textbf{Triangle}\\\textbf{(LiveJournal)}\\\textbf{undirected}}}
& \multicolumn{2}{c}{\makecell{\textbf{Triangle}\\\textbf{(Orkut)}\\\textbf{undirected}}}
& \makecell{\textbf{Andersen}\\\textbf{(Large)}}
& \makecell{\textbf{Andersen}\\\textbf{(Medium)}}
& \makecell{\textbf{TC}\\\textbf{(USRoad)}}
& \makecell{\textbf{SG}\\\textbf{(G10)}} \\
{\scriptsize\textbf{Input edges}} & \multicolumn{2}{c}{{\scriptsize 86,220,856 + 10 seeds}} & \multicolumn{2}{c}{{\scriptsize 234,370,166 + 10 seeds}} & \multicolumn{2}{c}{{\scriptsize 86,220,856}} & \multicolumn{2}{c}{{\scriptsize 234,370,166}} & {\scriptsize 67,999,999} & {\scriptsize 33,999,999} & {\scriptsize 165,435} & {\scriptsize 99,805} \\
{\scriptsize\textbf{Output edges}} & \multicolumn{2}{c}{{\scriptsize 3,727,171}} & \multicolumn{2}{c}{{\scriptsize 2,936,543}} & \multicolumn{2}{c}{{\scriptsize 1,793,261,106}} & \multicolumn{2}{c}{{\scriptsize 3,765,505,086}} & {\scriptsize 5,315,019} & {\scriptsize 2,654,871} & {\scriptsize 871,365,688} & {\scriptsize 100,000,000} \\
\cmidrule(lr){2-3}
\cmidrule(lr){4-5}
\cmidrule(lr){6-7}
\cmidrule(lr){8-9}
\textbf{Nproc} & \textbf{3D} & \textbf{2D}
& \textbf{3D} & \textbf{2D}
& \textbf{3D} & \textbf{2D}
& \textbf{3D} & \textbf{2D}
& \textbf{2D}
& \textbf{2D}
& \textbf{2D}
& \textbf{2D} \\
\midrule
32   & 24.071 & 16.505 & 76.296 & 48.494 & 27.088 & 25.579 & 98.763 & 88.609 & 14.314 & 7.322 & 18.467 & 68.212 \\
64   & 30.334 & 15.147 & 87.942 & 44.806 & 21.681 & 18.666 & 78.765 & 68.361 & 12.586 & 6.378 & 11.870 & 51.237 \\
128  & 16.290 & 9.971  & 45.375 & 27.917 & 11.906 & 12.467 & 42.005 & 38.555 & 10.489 & 4.923 & 6.606  & 31.070 \\
256  & 14.324 & 8.479  & 34.264 & 24.493 & 8.749  & 8.768  & 26.731 & 26.879 & 7.409  & 3.516 & 4.294  & 16.301 \\
512  & 15.995 & 6.562  & 43.051 & 16.578 & 8.090  & 6.654  & 23.091 & 19.229 & 6.290  & 3.347 & 3.403  & 9.749  \\
1024 & 10.804 & 5.952  & 30.188 & 17.692 & 8.599  & 7.676  & 24.956 & 21.521 & 4.637  & 2.020 & 4.058  & 5.448  \\
2048 & 10.263 & 4.828  & 28.293 & 15.198 & 8.042  & 7.116  & 23.570 & 19.571 & 3.591  & 2.317 & 7.166  & 5.091  \\
\bottomrule
\end{tabular}
\end{table*}

\section{Evaluation}
\label{evaluation}

We evaluate our system on two leadership-class HPC systems, Polaris and Aurora. Each Polaris node contains a single 32-core AMD EPYC 7543P (``Milan'') CPU at 2.8\,GHz and 512\,GB of DDR4 memory. Each Aurora node contains two 52-core Intel Xeon Max (``Sapphire Rapids'') CPUs, 1\,TB of DDR5 memory, and 128\,GB of CPU-attached HBM.
On Polaris, we allocate 32 MPI ranks per node so that each rank is mapped to one physical core. On Aurora, we place at most 64 MPI ranks per node.
Our implementation is written in C++23 with MPICH and compiled with Clang 19. All reported results use the same runtime implementation, partitioning strategy, and measurement methodology.

We evaluate five queries: \textsc{Peer Influence}, \textsc{Triangle}, transitive closure (\textsc{TC}), same generation (\textsc{SG}), and Andersen-style context-sensitive points-to analysis. These queries cover a range of workload characteristics, including recursion (\textsc{TC}, \textsc{SG}, and Andersen), cyclic recursive multi-way joins (\textsc{Peer Influence}), cyclic non-recursive multi-way joins (\textsc{Triangle}), graph analytics (\textsc{TC} and \textsc{SG}), and join-intensive program analysis with substantial intermediate-result pressure (Andersen's points-to analysis). We run these queries on several large graphs: LiveJournal, Orkut, USRoad, and G10; points-to analysis is run on two synthetic input programs, \textsc{Medium} and \textsc{Large}.

We compare our system against several well-known single-node and multi-node systems. For single-node systems, we use
 Souffl\'e~\cite{jordan2016souffle}, FlowLog~\cite{zhao2025flowlog}, and DuckDB~\cite{raasveldt2019duckdb}. For multi-node Datalog engines, we use Slog~\cite{gilray2024datalog} and Multi-Node Multi-GPU Datalog~\cite{shovon2025multi}.
We measure the end-to-end runtime of our system after MPI initialization; this runtime includes reading input data, redistribution, replication, index construction, iterative evaluation, and materialization of results. For distributed runs, we report the maximum runtime across all ranks, and we separately report communication time for selected workloads.

\subsection{Scalability under Shared-Nothing Execution}
\label{sec:scalability-shared-nothing}

Tables~\ref{tab:polaris-2d-3d} and~\ref{tab:aurora-2d-3d} report the end-to-end
runtime from 32 to 2,048 processes on Polaris and Aurora.
On both systems, despite the use of communication-intensive collectives such as all-to-all and all-gather, our system achieves good strong scaling. The gains, however, are not uniform across workloads: most benchmarks scale well through 256--512 processes, after which improvements diminish. %
The degree of scalability depends on both the workload and the partitioning layout. On Polaris,
the runtime of Andersen (Large) decreases from 17.519\,s at 32 processes to 3.804\,s at
512 processes and 3.141\,s at 2,048 processes; on Aurora, it decreases from
14.314\,s to 6.290\,s and 3.591\,s. \textsc{Peer Influence} (Orkut) shows the same
overall trend, but with a stronger layout effect. In 3D, its runtime on Polaris
decreases from 68.247\,s at 32 processes to 39.556\,s at 512 processes and
23.317\,s at 2,048 processes. The 2D layout is faster when feasible,
decreasing from 30.765\,s at 64 processes to 10.087\,s at 512 processes and
7.955\,s at 2,048 processes, but it runs out of memory at 32 processes. Aurora
shows the same pattern: 3D decreases from 76.296\,s to 43.051\,s and
28.293\,s, while 2D decreases from 48.494\,s to 16.578\,s and 15.198\,s.

\begin{figure}[t]
    \centering
    \begin{subfigure}[t]{\columnwidth}
        \centering
        \includegraphics[width=\linewidth]{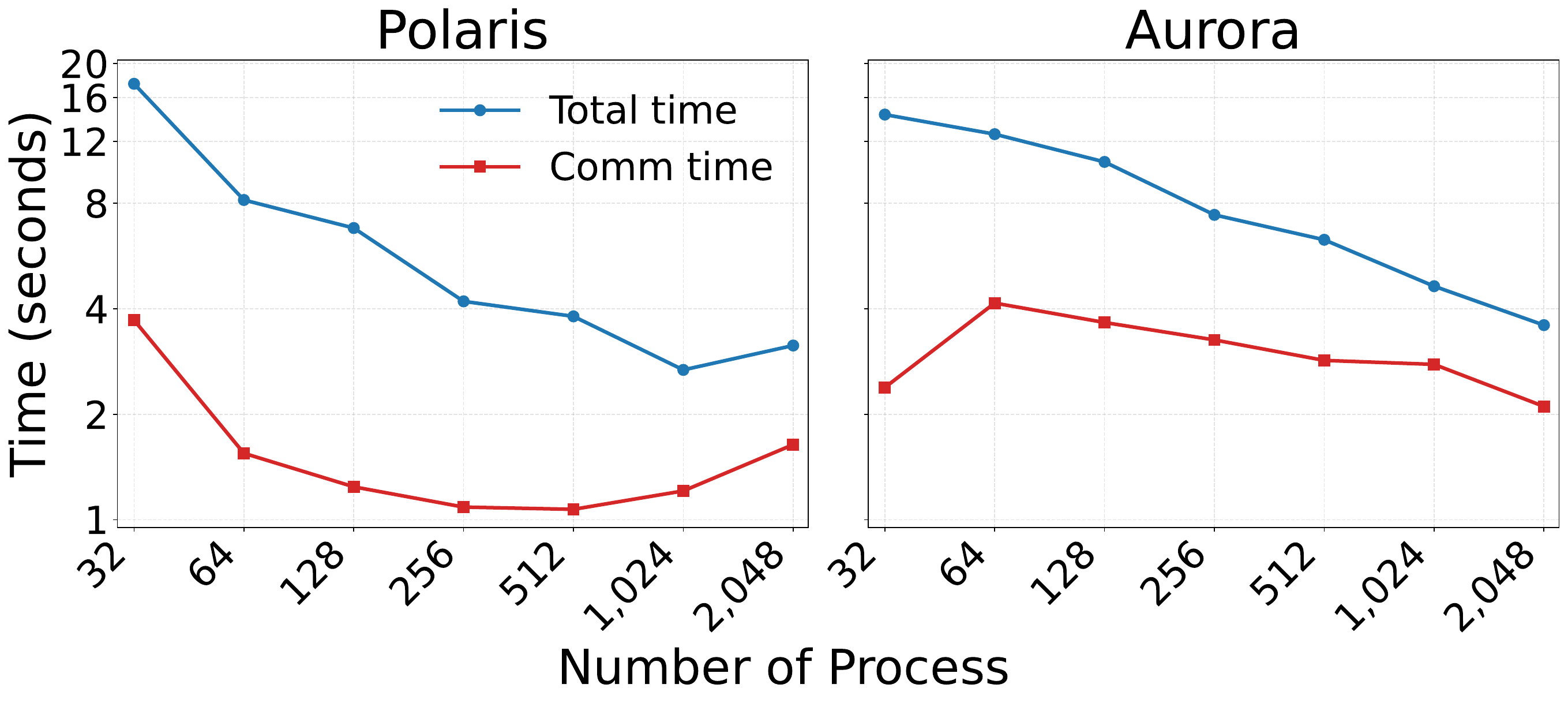}
        \caption{End-to-end runtime of Andersen (Large) on Polaris and Aurora.}
        \label{fig:andersen-large-scaling}
    \end{subfigure}

    \vspace{0.5em}
    \begin{subfigure}[t]{\columnwidth}
        \centering
        \includegraphics[width=0.73\linewidth]{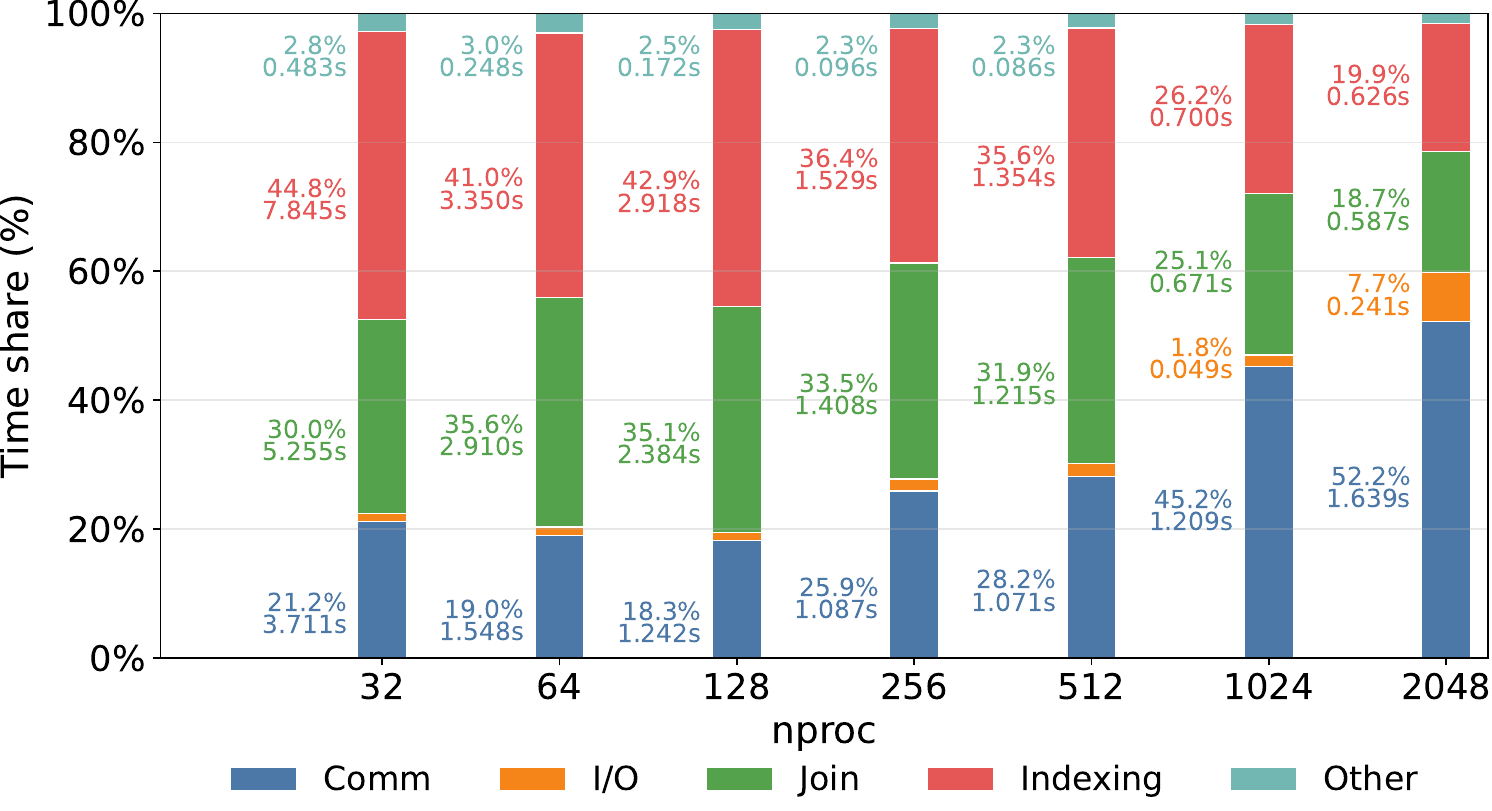}
        \caption{Runtime breakdown on Polaris.}
        \label{fig:andersen-large-breakdown-polaris}
    \end{subfigure}

    \vspace{0.5em}

    \begin{subfigure}[t]{\columnwidth}
        \centering
        \includegraphics[width=0.73\linewidth]{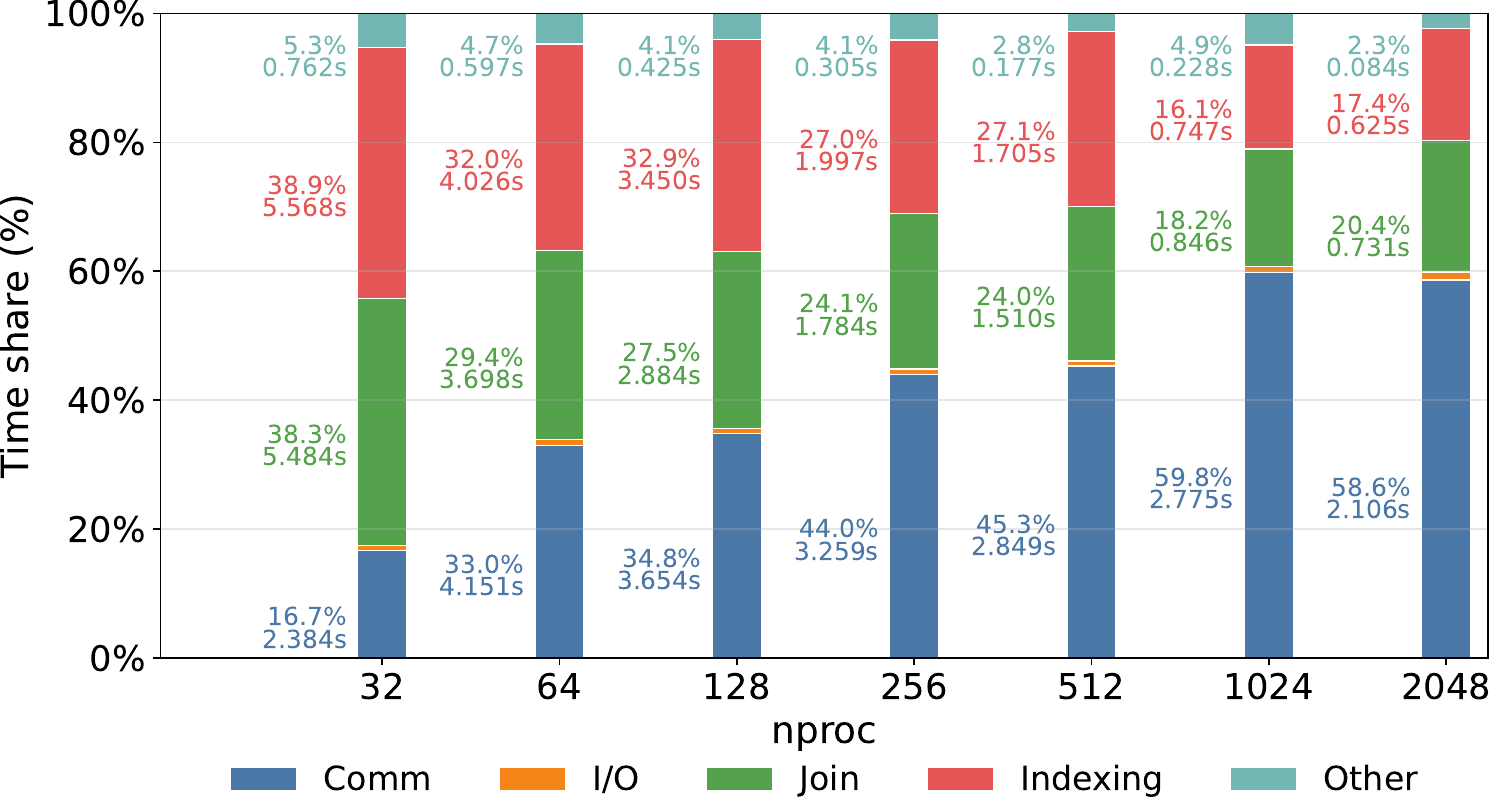}
        \caption{Runtime breakdown on Aurora.}
        \label{fig:andersen-large-breakdown-aurora}
    \end{subfigure}

\caption{Andersen (Large) under shared-nothing execution. The top panel shows
end-to-end runtime on Polaris and Aurora. The lower two panels show the runtime
breakdown.}
    \label{fig:andersen-large-combined}
    \vspace{-0.5cm}
\end{figure}
Several distinct cross-workload patterns emerge from these tables. First, most
benchmarks achieve their largest runtime reduction between 32 and 512
processes, indicating that the shared-nothing execution model is most effective while each rank still performs a substantial amount of local work. Second, scaling
regularity differs by workload. Andersen scales most smoothly on
both systems, while \textsc{Peer Influence} scales less regularly. \textsc{Triangle} and
\textsc{SG} continue to improve at larger process counts, but with diminishing returns.
\textsc{TC} shows that overly aggressive scaling can eventually degrade performance: its runtime increases beyond 512 processes. Third, these scaling patterns are consistent across Polaris and Aurora, suggesting that they are intrinsic to the workloads rather than to the platforms. Overall, these results show that our system achieves robust strong scaling across diverse workloads, although
the regularity and ultimate benefit of scaling depend on both the workload and the partitioning configuration.

To better illustrate scalability, consider Andersen (Large) in Figure~\ref{fig:andersen-large-combined}.
As the process count increases, our distributed
WCOJ execution shifts from being compute-bound to being communication-bound. In the top panel, runtime
drops steadily on both Polaris and Aurora as the number of processes
increases, showing that partitioning the workload across more ranks is
initially effective at reducing local computation. The breakdown plots show why
this improvement becomes less pronounced at larger scales. At small process
counts, join and indexing account for a large fraction of the runtime, indicating that
local join evaluation is the main bottleneck. As the process count grows, these
fractions shrink, while communication takes an increasing share of the total
runtime. On Polaris, the communication share increases from 21.2\% (32 processes) to
52.2\% (2,048 processes); on Aurora, it increases from 16.7\% to 58.6\%. This
shift indicates that, although local work continues to decrease with more processes, end-to-end
performance becomes increasingly constrained by communication and coordination overhead.

Tables~\ref{tab:polaris-2d-3d} and~\ref{tab:aurora-2d-3d} also show the trade-off
between 2D and 3D partitioning. When both layouts are feasible, 2D is faster.
On Polaris, the runtime of \textsc{Peer Influence} (Orkut) decreases from 30.765\,s at
64 processes to 7.955\,s at 2,048 processes in 2D, compared with 68.945\,s to
23.317\,s in 3D. \textsc{Triangle} (Orkut) shows the same pattern. However, 2D runs out
of memory for \textsc{Peer Influence} (Orkut) and \textsc{Triangle} (Orkut) at 32 processes on
Polaris, while 3D remains feasible. On Aurora, these 2D configurations remain feasible at 32 processes only because a single Aurora node has twice the memory capacity of a Polaris node.

\begin{table}[t]
\centering
\small
\caption{Communication share (\%) of end-to-end runtime for
\textsc{Peer Influence} (Orkut) under 2D and 3D partitioning.}
\label{tab:peer-comm-share}
\setlength{\tabcolsep}{5pt}
\begin{tabular}{lcccc}
\toprule
\textbf{Configuration} & \textbf{32} & \textbf{128} & \textbf{512} &\textbf{2048} \\
\midrule
Polaris 2D & OOM  & 25.1& 40.6&63.9 \\
Polaris 3D & 35.7 & 54.4& 79.1&86.8 \\
Aurora 2D  & 24.4 & 41.3& 58.8&78.1 \\
Aurora 3D  & 43.9 & 55.4& 78.3&88.8 \\
\bottomrule
\end{tabular}
\end{table}

Table~\ref{tab:peer-comm-share} shows that communication accounts for a larger share of the runtime under 3D partitioning than under 2D. On Polaris, the
communication share of \textsc{Peer Influence} grows from 26.5\% at 64 processes to
63.9\% at 2,048 processes in 2D, compared with 35.7\% at 32 processes to
86.8\% in 3D. On Aurora, it grows from 24.4\% to 78.1\% in 2D, compared with
43.9\% to 88.8\% in 3D. The runtimes in Tables~\ref{tab:polaris-2d-3d} and~\ref{tab:aurora-2d-3d} are consistent with this difference: 2D achieves lower
runtime when memory permits, while 3D incurs a heavier communication penalty. The more
complex 3D partitioning pattern may also make balanced execution harder to
maintain for some workloads. Overall, 2D favors runtime, while 3D favors memory feasibility on larger
inputs.

\subsection{Comparison with Existing Systems}
We next compare our system against existing single-node and distributed baselines to contextualize the practical performance of our distributed WCOJ architecture.
Table~\ref{tab:polaris-single-node} presents the single-node comparison on one Polaris node with 32-way parallelism. Overall, our system is consistently competitive and often faster on recursive and multi-way workloads. It achieves the best runtime on \textsc{Peer} (LJ), \textsc{Peer} (Orkut), \textsc{Triangle} (LJ), \textsc{Triangle} (Orkut), and \textsc{TC} (USRoad). FlowLog performs particularly well on the Andersen and \textsc{SG} (G10) benchmarks. These workloads require many iterations but produce small deltas per iteration, a pattern well suited to FlowLog's LSM-tree backend, which is tuned for fine-grained incremental updates.
DuckDB is the closest competitor on the two \textsc{Triangle} workloads, but it times out on \textsc{Peer} (Orkut) and produces incorrect results on the mutually recursive Andersen analyses, which it does not natively support. Souffl\'e, the long-standing gold standard, supports the richest set of Datalog language features, but it is slower than the newer engines on most workloads. Overall, the single-node results show that our design is competitive even without multi-node scaling.

\begin{table}[t]
\centering
\small
\setlength{\tabcolsep}{3.5pt}
\caption{Runtime (seconds) comparison on a single Polaris node with 32-way parallelism (r: MPI ranks; t: threads). Lower is better. \timeout: timeout (more than 500 seconds); \failed: failed or stopped before completion; \incorrect: incorrect result.}
\label{tab:polaris-single-node}
\begin{tabular}{lcccc}
\toprule
\textbf{Benchmark} & \makecell{\textbf{Ours}\\\textbf{(32r)}} & \makecell{\textbf{Souffl\'e}\\\textbf{(32t)}} & \makecell{\textbf{FlowLog}\\\textbf{(32t)}} & \makecell{\textbf{DuckDB}\\\textbf{(32t)}} \\
\midrule
Peer (LJ)     & 23.896                         & 48.322                         & 165.661                        & 117.821                        \\
Peer (Orkut)  & 68.247                         & 479.150                        & \failed                        & \timeout                       \\
Anders. (L)   & 17.519                         & 60.094                         & 9.408                          & 2.668\incorrect                \\
Anders. (M)   & 7.191                          & 26.510                         & 5.555                          & 1.664\incorrect                \\
Tri. (LJ)     & 28.516                         & 276.293                        & 381.262                        & 34.067                         \\
Tri. (Orkut)  & 91.546                         & \timeout                       & \failed                        & 161.513                        \\
TC (USRoad)   & 20.199                         & 190.276                        & 68.595                         & 146.823                        \\
SG (G10)      & 78.235                         & 57.347                         & 31.905                         & 468.424                        \\
\bottomrule
\end{tabular}
\vspace{-0.2cm}
\end{table}

In addition to single-node systems, we compare our system with existing distributed baselines. Table~\ref{tab:tc-distributed} reports the comparison with prior distributed systems for \textsc{TC} (USRoad) on Polaris. We focus on transitive closure here because it is one of the few workloads for which directly comparable results from prior distributed systems are available.
Across all process counts in Table~\ref{tab:tc-distributed}, our system outperforms Slog by a large margin (8.9--11.2$\times$). It also outperforms the multi-node multi-GPU Datalog engine at every process count, although the advantage narrows from 3.7$\times$ at 32 processes to 1.5$\times$ at 512 processes.
We interpret this result conservatively: it indicates that our shared-nothing CPU design is highly effective for distributed recursive evaluation, while accelerator-based systems may become more competitive at larger scales. Taken together, these results show that our approach is competitive on a single node and compares favorably with existing distributed systems at scale.

\section{Related Work}
\label{sec:related}

\emph{\textbf{Distributed Worst-Case Optimal Joins.}}
The HyperCube (Shares) algorithm~\cite{afrati2010optimizing,beame2017communication} evaluates a multi-way join in a single communication round by hashing each relation along the grid dimensions of the variables it contains and replicating it along the remaining dimensions; its communication cost has been studied extensively in the massively parallel computation model~\cite{koutris2018algorithmic}. Chu et al.~\cite{chu2015theory} pair HyperCube shuffling with Tributary Join, a variant of Leapfrog Triejoin, in the Myria parallel database, and ADJ~\cite{zhang2021fast} and EdgeFrame~\cite{fuchs2020edgeframe} bring WCOJ to Spark. BiGJoin~\cite{ammar2018distributed} instead distributes Generic Join as a Timely Dataflow computation, achieving worst-case optimal communication for subgraph queries. All of these systems target one-shot conjunctive or subgraph queries. Myria also supports recursive Datalog, but with binary joins~\cite{wang2015asynchronous}, and hypercube-style distribution policies for recursive Datalog have been analyzed theoretically~\cite{ketsman2018distribution}. To our knowledge, no prior system combines WCOJ with recursive fixpoint evaluation on distributed memory. Our work realizes hypercube partitioning with MPI subcube communicators and repartitions and re-replicates newly derived deltas in every iteration.

\emph{\textbf{GPU-Based Datalog Engines.}}
The massive memory bandwidth of GPUs has long been underexploited for
Datalog evaluation. Early attempts at running full Datalog on
GPUs~\cite{martinez2013datalog,wu2014red,shovon2022accelerating} failed
to outperform state-of-the-art CPU engines such as
Souffl\'{e}~\cite{scholz2016fast}. GPUlog~\cite{sun2025optimizing}
achieved the first order-of-magnitude speedup over the fastest CPU
engines on data-center-grade GPUs, introducing a hash-indexed sorted
array (HISA) index structure with virtualized trie access. Its
follow-up, VFLog~\cite{sun2025column}, demonstrated further
acceleration through a column-oriented data layout, and
SRDatalog~\cite{sun2026scaling} brings WCOJ to Datalog on a single
GPU. Our work does not
use hardware accelerators; however, our data partitioning mechanism is
orthogonal to on-node join acceleration. We believe combining our
distributed WCOJ framework with recent work on scaling GPU Datalog to
multiple nodes~\cite{shovon2025multi} is a promising direction for
future research.

\begin{table}[t]
\centering
\small
\caption{Runtime (seconds) comparison with existing distributed systems for \textsc{TC} (USRoad) on Polaris. Lower is better.}
\label{tab:tc-distributed}
\begin{tabular}{lccc}
\toprule
\textbf{Nproc} & \textbf{Ours} & \textbf{Slog} & \shortstack{\textbf{Multi-Node}\\\textbf{Multi-GPU}} \\
\midrule
32  & 20.20 & 180.05 & 75.07 \\
64  & 10.34 & 100.20 & 36.92 \\
128 & 6.04  & 58.28  & 18.06 \\
256 & 3.60  & 40.49  & 8.35 \\
512 & 2.50  & 24.75  & 3.67 \\
\bottomrule
\end{tabular}
\vspace{-0.3cm}
\end{table}
\emph{\textbf{Incremental Datalog Processing.}}
While semi-na\"ive evaluation optimizes a single fixpoint computation, evolving data requires incremental maintenance. Frameworks such as Differential Dataflow~\cite{mcsherry2013differential} and DBSP~\cite{dbsp} track state changes algebraically to propagate insertions and deletions efficiently. Systems such as Differential Datalog~\cite{ryzhyk2019differential} and FlowLog~\cite{zhao2025flowlog} leverage these principles to achieve high-performance incremental evaluation. Unlike these incremental engines, our work targets distributed batch processing, the dominant execution model for established Datalog applications such as program analysis. Extending our distributed architecture to support incremental streaming workloads remains future work.

\emph{\textbf{Robust Join Processing.}}
To address intermediate blowup in binary joins, recent systems introduce robust join processing techniques beyond WCOJ. Approaches such as Diamond Hardened Joins~\cite{birler2024robust} and TreeTracker Join~\cite{treetracker} enable early tuple pruning and backjumping, achieving worst-case optimal bounds for acyclic queries. Similarly, Lookahead Information Passing~\cite{zhu2017looking} and Robust Predicate Transfer~\cite{debuking} use Bloom filters to approximate semijoin reductions and pre-filter join inputs, and a distributed variant of predicate transfer~\cite{yang2025accelerate} applies this idea to reduce inter-node data exchange. However, these techniques predominantly target acyclic queries and do not easily extend to the cyclic joins prevalent in Datalog workloads such as program analysis. They are therefore complementary to our WCOJ-based approach, which also handles cyclic queries.

\section{Conclusion and Future Work}
\label{sec:conclusion}

We presented, to our knowledge, the first distributed-memory Datalog engine that uses Worst-Case Optimal Joins for recursive fixpoint evaluation on shared-nothing HPC systems. To address the fundamental limitations of chained binary joins in distributed recursive evaluation, we extended HyperCube partitioning from a single shuffle to iterative execution: each semi-na\"ive iteration repartitions and re-replicates only the newly derived tuples with MPI collectives over subcube communicators, while reusing the same layout so that all WCOJ intersections remain rank-local. On graph pattern mining and program analysis benchmarks on Polaris and Aurora, our system achieves up to $14.3\times$ speedup when scaling from 32 to 2,048 processes, is faster than established single-node engines on most benchmarks, and outperforms existing distributed Datalog engines on transitive closure. Our analysis of partition dimensionality further shows that 2D partitioning minimizes runtime, whereas 3D partitioning remains feasible where 2D runs out of memory. These results confirm that worst-case optimal join execution is practically viable at HPC scale, establishing a new foundation for scalable graph analytics, program analysis, and neuro-symbolic AI.

Our evaluation also shows that communication becomes the dominant cost at large process counts: at 2,048 processes, it accounts for 52--59\% of the runtime of Andersen (Large) and 87--89\% of the runtime of \textsc{Peer Influence} (Orkut) under 3D partitioning. Every iteration of our pipeline redistributes newly derived tuples through a non-uniform all-to-all exchange whose message sizes vary widely across ranks and across iterations, and our current implementation relies on the MPI library's default \texttt{MPI\_Alltoallv}. A natural next step is therefore to replace this exchange with optimized non-uniform all-to-all algorithms, such as the Bruck-based and parameterized algorithms of Fan et al.~\cite{fan2022optimizing,fan2025parameterized}, which have been shown to outperform standard \texttt{MPI\_Alltoallv} implementations on non-uniform workloads. Because the best algorithm and its parameters depend on the message-size distribution, the process count, and the machine, we also plan to integrate the machine-learning-driven auto-tuner of Qi et al.~\cite{qi2025machine}, which selects the \texttt{MPI\_Alltoallv} algorithm and its parameters automatically. Such tuning is particularly attractive for iterative evaluation, where the size of the delta relation, and therefore the communication pattern, changes from one iteration to the next. Other directions include cost-based selection of partition keys and hypercube dimensionality, skew-aware adaptive replication, and combining our partitioning scheme with GPU-accelerated local joins.

\bibliographystyle{IEEEtran}
\bibliography{cite}

@String{JACM = "J. ACM" }

@String{Computing = "Computing" }

@String{Computer = "{IEEE} Computer" }

@String{Springer = "Springer-Verlag" }

@article{wang2023free,
author = {Wang, Yisu Remy and Willsey, Max and Suciu, Dan},
title = {{Free Join}: Unifying Worst-Case Optimal and Traditional Joins},
year = {2023},
issue_date = {June 2023},
publisher = {Association for Computing Machinery},
address = {New York, NY, USA},
volume = {1},
number = {2},
url = {https://doi.org/10.1145/3589295},
doi = {10.1145/3589295},
journal = {Proc. ACM Manag. Data},
month = jun,
articleno = {150},
numpages = {23}
}

@inproceedings{veldhuizen2014leapfrog,
  title={{Leapfrog Triejoin}: A simple, worst-case optimal join algorithm},
  author={Veldhuizen, Todd L},
  booktitle={Proc. International Conference on Database Theory},
  year={2014},
  url = {https://www.openproceedings.org/2014/conf/icdt/Veldhuizen14.pdf}
}

@article{ngo2018worst,
author = {Ngo, Hung Q. and Porat, Ely and R\'{e}, Christopher and Rudra, Atri},
title = {Worst-case Optimal Join Algorithms},
year = {2018},
issue_date = {June 2018},
publisher = {Association for Computing Machinery},
address = {New York, NY, USA},
volume = {65},
number = {3},
issn = {0004-5411},
url = {https://doi.org/10.1145/3180143},
doi = {10.1145/3180143},
journal = {J. ACM},
month = mar,
articleno = {16},
numpages = {40}
}

@article{10.14778/3407790.3407797,
author = {Freitag, Michael and Bandle, Maximilian and Schmidt, Tobias and Kemper, Alfons and Neumann, Thomas},
title = {Adopting worst-case optimal joins in relational database systems},
year = {2020},
issue_date = {August 2020},
publisher = {VLDB Endowment},
volume = {13},
number = {12},
issn = {2150-8097},
url = {https://doi.org/10.14778/3407790.3407797},
doi = {10.14778/3407790.3407797},
journal = {Proc. VLDB Endow.},
month = jul,
pages = {1891–1904},
numpages = {14}
}

@inproceedings{raasveldt2019duckdb,
author = {Raasveldt, Mark and M\"{u}hleisen, Hannes},
title = {{DuckDB}: an Embeddable Analytical Database},
year = {2019},
isbn = {9781450356435},
publisher = {Association for Computing Machinery},
address = {New York, NY, USA},
url = {https://doi.org/10.1145/3299869.3320212},
doi = {10.1145/3299869.3320212},
booktitle = {Proceedings of the 2019 International Conference on Management of Data},
pages = {1981–1984},
numpages = {4},
location = {Amsterdam, Netherlands},
series = {SIGMOD '19}
}

@article{atserias2013size,
  title={Size bounds and query plans for relational joins},
  author={Atserias, Albert and Grohe, Martin and Marx, D{\'a}niel},
  journal={SIAM Journal on Computing},
  volume={42},
  number={4},
  pages={1737--1767},
  year={2013},
  url = {https://doi.org/10.1137/110859440},
  doi = {10.1137/110859440},
  publisher={SIAM}
}

@article{beame2017communication,
  title={Communication steps for parallel query processing},
  author={Beame, Paul and Koutris, Paraschos and Suciu, Dan},
  journal={Journal of the ACM (JACM)},
  volume={64},
  number={6},
  pages={1--58},
  year={2017},
  publisher={ACM New York, NY, USA},
  doi={10.1145/3125644}
}

@article{birler2024robust,
author = {Birler, Altan and Kemper, Alfons and Neumann, Thomas},
title = {Robust Join Processing with {Diamond Hardened Joins}},
year = {2024},
issue_date = {July 2024},
publisher = {VLDB Endowment},
volume = {17},
number = {11},
issn = {2150-8097},
url = {https://doi.org/10.14778/3681954.3681995},
doi = {10.14778/3681954.3681995},
journal = {Proc. VLDB Endow.},
month = jul,
pages = {3215–3228},
numpages = {14}
}

@article{treetracker,
author = {Hu, Zeyuan and Wang, Yisu Remy and Miranker, Daniel P},
title = {{TreeTracker} Join: Simple, Optimal, Fast},
year = {2025},
publisher = {Association for Computing Machinery},
address = {New York, NY, USA},
issn = {0362-5915},
url = {https://doi.org/10.1145/3774325},
doi = {10.1145/3774325},
note = {Just Accepted},
journal = {ACM Trans. Database Syst.},
month = oct
}

@article{debuking,
author = {Zhao, Junyi and Su, Kai and Yang, Yifei and Yu, Xiangyao and Koutris, Paraschos and Zhang, Huanchen},
title = {Debunking the Myth of Join Ordering: Toward Robust {SQL} Analytics},
year = {2025},
issue_date = {June 2025},
publisher = {Association for Computing Machinery},
address = {New York, NY, USA},
volume = {3},
number = {3},
url = {https://doi.org/10.1145/3725283},
doi = {10.1145/3725283},
journal = {Proc. ACM Manag. Data},
month = jun,
articleno = {146},
numpages = {28}

}

@inproceedings{yannakakis1981algorithms,
author = {Yannakakis, Mihalis},
title = {Algorithms for acyclic database schemes},
year = {1981},
publisher = {VLDB Endowment},
booktitle = {Proceedings of the Seventh International Conference on Very Large Data Bases - Volume 7},
pages = {82–94},
numpages = {13},
location = {Cannes, France},
series = {VLDB '81},
url = {https://dl.acm.org/doi/10.5555/1286831.1286840}
}

@book{abiteboul1995foundations,
  title     = {Foundations of Databases},
  author    = {Abiteboul, Serge and Hull, Richard and Vianu, Victor},
  volume    = {8},
  year      = {1995},
  publisher = {Addison-Wesley},
  address   = {Reading, MA}
}

@article{ceri1989you,
author = {Ceri, S. and Gottlob, G. and Tanca, L.},
title = {What You Always Wanted to Know About {Datalog} (And Never Dared to Ask)},
year = {1989},
issue_date = {March 1989},
publisher = {IEEE Educational Activities Department},
address = {USA},
volume = {1},
number = {1},
issn = {1041-4347},
url = {https://doi.org/10.1109/69.43410},
doi = {10.1109/69.43410},
journal = {IEEE Trans. on Knowl. and Data Eng.},
month = mar,
pages = {146–166},
numpages = {21}
}

@article{li2023scallop,
author = {Li, Ziyang and Huang, Jiani and Naik, Mayur},
title = {{Scallop}: A Language for Neurosymbolic Programming},
year = {2023},
issue_date = {June 2023},
publisher = {Association for Computing Machinery},
address = {New York, NY, USA},
volume = {7},
number = {PLDI},
url = {https://doi.org/10.1145/3591280},
doi = {10.1145/3591280},
journal = {Proc. ACM Program. Lang.},
month = jun,
articleno = {166},
numpages = {25}

}

@inproceedings{gilray2021compiling,
author = {Gilray, Thomas and Kumar, Sidharth and Micinski, Kristopher},
title = {Compiling data-parallel {Datalog}},
year = {2021},
isbn = {9781450383257},
publisher = {Association for Computing Machinery},
address = {New York, NY, USA},
url = {https://doi.org/10.1145/3446804.3446855},
doi = {10.1145/3446804.3446855},
booktitle = {Proceedings of the 30th ACM SIGPLAN International Conference on Compiler Construction},
pages = {23–35},
numpages = {13},
location = {Virtual, Republic of Korea},
series = {CC 2021}
}

@inproceedings{scholz2016fast,
author = {Scholz, Bernhard and Jordan, Herbert and Suboti\'{c}, Pavle and Westmann, Till},
title = {On fast large-scale program analysis in {Datalog}},
year = {2016},
isbn = {9781450342414},
publisher = {Association for Computing Machinery},
address = {New York, NY, USA},
url = {https://doi.org/10.1145/2892208.2892226},
doi = {10.1145/2892208.2892226},
booktitle = {Proceedings of the 25th International Conference on Compiler Construction},
pages = {196–206},
numpages = {11},
location = {Barcelona, Spain},
series = {CC '16}
}

@inproceedings{flores2020datalog,
author = {Flores-Montoya, Antonio and Schulte, Eric},
title = {{Datalog} disassembly},
year = {2020},
isbn = {978-1-939133-17-5},
publisher = {USENIX Association},
address = {USA},
booktitle = {Proceedings of the 29th USENIX Conference on Security Symposium},
articleno = {61},
numpages = {18},
series = {SEC'20},
url = {https://dl.acm.org/doi/10.5555/3489212.3489273}
}

@article{zhao2025flowlog,
author = {Zhao, Hangdong and Yu, Zhenghong and Rao, Srinag and Frisk, Simon and Fan, Zhiwei and Koutris, Paraschos},
title = {{FlowLog}: Efficient and Extensible {Datalog} via Incrementality},
year = {2025},
issue_date = {November 2025},
publisher = {VLDB Endowment},
volume = {19},
number = {3},
issn = {2150-8097},
url = {https://doi.org/10.14778/3778092.3778098},
doi = {10.14778/3778092.3778098},
journal = {Proc. VLDB Endow.},
month = nov,
pages = {361–374},
numpages = {14}
}

@inproceedings{jordan2016souffle,
author="Jordan, Herbert and Scholz, Bernhard and Suboti{\'{c}}, Pavle",
editor="Chaudhuri, Swarat and Farzan, Azadeh",
title="{Souffl{\'e}}: On Synthesis of Program Analyzers",
booktitle="Computer Aided Verification",
year="2016",
publisher="Springer International Publishing",
address="Cham",
pages="422--430",
url = "https://doi.org/10.1007/978-3-319-41540-6\_23"
}

@inproceedings{sun2025column,
author = {Sun, Yihao and Kumar, Sidharth and Gilray, Thomas and Micinski, Kristopher},
title = {Column-oriented {Datalog} on the {GPU}},
year = {2025},
isbn = {978-1-57735-897-8},
publisher = {AAAI Press},
url = {https://doi.org/10.1609/aaai.v39i14.33665},
doi = {10.1609/aaai.v39i14.33665},
booktitle = {Proceedings of the Thirty-Ninth AAAI Conference on Artificial Intelligence and Thirty-Seventh Conference on Innovative Applications of Artificial Intelligence and Fifteenth Symposium on Educational Advances in Artificial Intelligence},
articleno = {1692},
numpages = {9},
series = {AAAI'25/IAAI'25/EAAI'25}
}

@inproceedings{shovon2025multi,
author = {Shovon, Ahmedur Rahman and Sun, Yihao and Micinski, Kristopher and Gilray, Thomas and Kumar, Sidharth},
title = {Multi-Node Multi-{GPU} {Datalog}},
year = {2025},
isbn = {9798400715372},
publisher = {Association for Computing Machinery},
address = {New York, NY, USA},
url = {https://doi.org/10.1145/3721145.3730431},
doi = {10.1145/3721145.3730431},
booktitle = {Proceedings of the 39th ACM International Conference on Supercomputing},
pages = {822–836},
numpages = {15},
location = {
},
series = {ICS '25}
}

@inproceedings{shkapsky2016big,
author = {Shkapsky, Alexander and Yang, Mohan and Interlandi, Matteo and Chiu, Hsuan and Condie, Tyson and Zaniolo, Carlo},
title = {Big Data Analytics with {Datalog} Queries on {Spark}},
year = {2016},
isbn = {9781450335317},
publisher = {Association for Computing Machinery},
address = {New York, NY, USA},
url = {https://doi.org/10.1145/2882903.2915229},
doi = {10.1145/2882903.2915229},
booktitle = {Proceedings of the 2016 International Conference on Management of Data},
pages = {1135–1149},
numpages = {15},
location = {San Francisco, California, USA},
series = {SIGMOD '16}
}

@inproceedings{nenov2015rdfox,
author = {Nenov, Yavor and Piro, Robert and Motik, Boris and Horrocks, Ian and Wu, Zhe and Banerjee, Jay},
title = {{RDFox}: A Highly-Scalable {RDF} Store},
year = {2015},
isbn = {978-3-319-25009-0},
publisher = {Springer-Verlag},
address = {Berlin, Heidelberg},
url = {https://doi.org/10.1007/978-3-319-25010-6\_1},
doi = {10.1007/978-3-319-25010-6_1},
booktitle = {The Semantic Web - ISWC 2015: 14th International Semantic Web Conference, Bethlehem, PA, USA, October 11-15, 2015, Proceedings, Part II},
pages = {3–20},
numpages = {18},
location = {Bethlehem, PA, USA}
}

@inproceedings{gu2019rasql,
author = {Gu, Jiaqi and Watanabe, Yugo H. and Mazza, William A. and Shkapsky, Alexander and Yang, Mohan and Ding, Ling and Zaniolo, Carlo},
title = {{RaSQL}: Greater Power and Performance for Big Data Analytics with Recursive-aggregate-{SQL} on {Spark}},
year = {2019},
isbn = {9781450356435},
publisher = {Association for Computing Machinery},
address = {New York, NY, USA},
url = {https://doi.org/10.1145/3299869.3324959},
doi = {10.1145/3299869.3324959},
booktitle = {Proceedings of the 2019 International Conference on Management of Data},
pages = {467–484},
numpages = {18},
location = {Amsterdam, Netherlands},
series = {SIGMOD '19}
}

@inproceedings{sun2025optimizing,
author = {Sun, Yihao and Shovon, Ahmedur Rahman and Gilray, Thomas and Kumar, Sidharth and Micinski, Kristopher},
title = {Optimizing {Datalog} for the {GPU}},
year = {2025},
isbn = {9798400706981},
publisher = {Association for Computing Machinery},
address = {New York, NY, USA},
url = {https://doi.org/10.1145/3669940.3707274},
doi = {10.1145/3669940.3707274},
booktitle = {Proceedings of the 30th ACM International Conference on Architectural Support for Programming Languages and Operating Systems, Volume 1},
pages = {762–776},
numpages = {15},
location = {Rotterdam, Netherlands},
series = {ASPLOS '25}
}

@inproceedings{shovon2022accelerating,
  title={Accelerating {Datalog} applications with {cuDF}},
  author={Shovon, Ahmedur Rahman and Dyken, Landon Richard and Green, Oded and Gilray, Thomas and Kumar, Sidharth},
  booktitle={2022 IEEE/ACM Workshop on Irregular Applications: Architectures and Algorithms (IA3)},
  pages={41--45},
  year={2022},
  organization={IEEE},
  url={https://doi.org/10.1109/IA356718.2022.00012}
}

@article{seo2013socialite,
author = {Seo, Jiwon and Park, Jongsoo and Shin, Jaeho and Lam, Monica S.},
title = {Distributed {SociaLite}: a {Datalog}-based language for large-scale graph analysis},
year = {2013},
issue_date = {September 2013},
publisher = {VLDB Endowment},
volume = {6},
number = {14},
issn = {2150-8097},
url = {https://doi.org/10.14778/2556549.2556572},
doi = {10.14778/2556549.2556572},
journal = {Proc. VLDB Endow.},
month = sep,
pages = {1906–1917},
numpages = {12}
}

@article{gilray2024datalog,
author = {Gilray, Thomas and Sahebolamri, Arash and Sun, Yihao and Kunapaneni, Sowmith and Kumar, Sidharth and Micinski, Kristopher},
title = {{Datalog} with First-Class Facts},
year = {2024},
issue_date = {November 2024},
publisher = {VLDB Endowment},
volume = {18},
number = {3},
issn = {2150-8097},
url = {https://doi.org/10.14778/3712221.3712232},
doi = {10.14778/3712221.3712232},
journal = {Proc. VLDB Endow.},
month = nov,
pages = {651–665},
numpages = {15}
}

@article{balbin1987generalization,
author = {Balbin, I. and Ramamohanarao, K.},
title = {A generalization of the differential approach to recursive query evaluation},
year = {1987},
issue_date = {Sept. 1987},
publisher = {Elsevier Science Inc.},
address = {USA},
volume = {4},
number = {3},
issn = {0743-1066},
url = {https://doi.org/10.1016/0743-1066(87)90004-5},
doi = {10.1016/0743-1066(87)90004-5},
journal = {J. Log. Program.},
month = sep,
pages = {259–262},
numpages = {4}
}

@inproceedings{ceri1986translation,
  title={Translation and optimization of logic queries: The algebraic approach},
  author={Ceri, Stefano and Gottlob, Georg and Lavazza, LUIGI ANTONIO and others},
  booktitle={Proceedings of the 12th International Conference on Very Large Data Bases},
  pages={395--402},
  year={1986},
  organization={Morgan Kaufmann Publishers Inc.},
  url = {https://dl.acm.org/doi/10.5555/645913.671468}
}

@inproceedings{mcsherry2013differential,
  title={{Differential Dataflow}},
  author={McSherry, Frank and Murray, Derek Gordon and Isaacs, Rebecca and Isard, Michael},
  booktitle={CIDR},
  year={2013},
  url={https://www.cidrdb.org/cidr2013/Papers/CIDR13\_Paper111.pdf}
}

@article{ryzhyk2019differential,
  title={Differential {Datalog}},
  author={Ryzhyk, Leonid and Budiu, Mihai},
  journal={Datalog 2.0},
  volume={2},
  pages={4--5},
  year={2019},
  url={https://doi.org/10.31224/3140}
}

@article{dbsp,
author = {Budiu, Mihai and Chajed, Tej and McSherry, Frank and Ryzhyk, Leonid and Tannen, Val},
title = {{DBSP}: Automatic Incremental View Maintenance for Rich Query Languages},
year = {2023},
issue_date = {March 2023},
publisher = {VLDB Endowment},
volume = {16},
number = {7},
issn = {2150-8097},
url = {https://doi.org/10.14778/3587136.3587137},
doi = {10.14778/3587136.3587137},
journal = {Proc. VLDB Endow.},
month = mar,
pages = {1601–1614},
numpages = {14}
}

@inproceedings{wu2014red,
author = {Wu, Haicheng and Diamos, Gregory and Sheard, Tim and Aref, Molham and Baxter, Sean and Garland, Michael and Yalamanchili, Sudhakar},
title = {{Red Fox}: An Execution Environment for Relational Query Processing on {GPUs}},
year = {2018},
isbn = {9781450326704},
publisher = {Association for Computing Machinery},
address = {New York, NY, USA},
url = {https://doi.org/10.1145/2544137.2544166},
doi = {10.1145/2544137.2544166},
booktitle = {Proceedings of Annual IEEE/ACM International Symposium on Code Generation and Optimization},
pages = {44–54},
numpages = {11},
location = {Orlando, FL, USA},
series = {CGO '14}
}

@inproceedings{martinez2013datalog,
  title={A {Datalog} engine for {GPUs}},
  author={Mart{\'\i}nez-Angeles, Carlos Alberto and Dutra, In{\^e}s and Costa, V{\'\i}tor Santos and Buenabad-Ch{\'a}vez, Jorge},
  booktitle={International Conference on Applications of Declarative Programming and Knowledge Management},
  pages={152--168},
  year={2013},
  organization={Springer},
  url={https://doi.org/10.1007/978-3-319-08909-6\_10}
}

@article{smaragdakis2015pointer,
author = {Smaragdakis, Yannis and Balatsouras, George},
title = {Pointer Analysis},
year = {2015},
issue_date = {4 2015},
publisher = {Now Publishers Inc.},
address = {Hanover, MA, USA},
volume = {2},
number = {1},
issn = {2325-1107},
url = {https://doi.org/10.1561/2500000014},
doi = {10.1561/2500000014},
journal = {Found. Trends Program. Lang.},
month = apr,
pages = {1–69},
numpages = {69}
}

@article{wu2025honeycomb,
author = {Wu, Jiacheng and Suciu, Dan},
title = {{HoneyComb}: A Parallel Worst-Case Optimal Join on Multicores},
year = {2025},
issue_date = {June 2025},
publisher = {Association for Computing Machinery},
address = {New York, NY, USA},
volume = {3},
number = {3},
url = {https://doi.org/10.1145/3725307},
doi = {10.1145/3725307},
journal = {Proc. ACM Manag. Data},
month = jun,
articleno = {170},
numpages = {27}
}

@inproceedings{fan2022optimizing,
  author    = {Fan, Ke and Gilray, Thomas and Pascucci, Valerio and Huang, Xuan and Micinski, Kristopher and Kumar, Sidharth},
  title     = {Optimizing the {B}ruck Algorithm for Non-uniform All-to-all Communication},
  booktitle = {Proceedings of the 31st International Symposium on High-Performance Parallel and Distributed Computing},
  series    = {HPDC '22},
  pages     = {172--184},
  publisher = {ACM},
  address   = {Minneapolis, MN, USA},
  year      = {2022},
  doi       = {10.1145/3502181.3531468}
}

@inproceedings{fan2025parameterized,
  author    = {Fan, Ke and Domke, Jens and Ba, Seydou and Kumar, Sidharth},
  title     = {Parameterized Algorithms for Non-uniform All-to-all},
  booktitle = {Proceedings of the 34th International Symposium on High-Performance Parallel and Distributed Computing},
  series    = {HPDC '25},
  publisher = {ACM},
  address   = {Notre Dame, IN, USA},
  year      = {2025},
  doi       = {10.1145/3731545.3731590}
}

@inproceedings{qi2025machine,
  author    = {Qi, Kunting and Fan, Ke and Domke, Jens and Ba, Seydou and Vishwanath, Venkatram and Papka, Michael E. and Kumar, Sidharth},
  title     = {Machine Learning Driven Auto-Tuning for Non-Uniform All-to-All Collectives},
  booktitle = {2025 IEEE 32nd International Conference on High Performance Computing, Data, and Analytics (HiPC)},
  pages     = {151--160},
  publisher = {IEEE},
  address   = {Hyderabad, India},
  year      = {2025},
  doi       = {10.1109/HiPC66333.2025.00024}
}

@inproceedings{afrati2010optimizing,
  author    = {Afrati, Foto N. and Ullman, Jeffrey D.},
  title     = {Optimizing Joins in a Map-Reduce Environment},
  booktitle = {Proceedings of the 13th International Conference on Extending Database Technology (EDBT)},
  pages     = {99--110},
  year      = {2010},
  publisher = {ACM},
  doi       = {10.1145/1739041.1739056}
}

@inproceedings{chu2015theory,
  author    = {Chu, Shumo and Balazinska, Magdalena and Suciu, Dan},
  title     = {From Theory to Practice: Efficient Join Query Evaluation in a Parallel Database System},
  booktitle = {Proceedings of the 2015 ACM SIGMOD International Conference on Management of Data},
  pages     = {63--78},
  year      = {2015},
  publisher = {ACM},
  doi       = {10.1145/2723372.2750545}
}

@article{ammar2018distributed,
  author    = {Ammar, Khaled and McSherry, Frank and Salihoglu, Semih and Joglekar, Manas},
  title     = {Distributed Evaluation of Subgraph Queries Using Worst-Case Optimal Low-Memory Dataflows},
  journal   = {Proc. VLDB Endow.},
  volume    = {11},
  number    = {6},
  pages     = {691--704},
  year      = {2018},
  doi       = {10.14778/3184470.3184473}
}

@inproceedings{zhang2021fast,
  author    = {Zhang, Hao and Qiao, Miao and Yu, Jeffrey Xu and Cheng, Hong},
  title     = {Fast Distributed Complex Join Processing},
  booktitle = {2021 IEEE 37th International Conference on Data Engineering (ICDE)},
  pages     = {2087--2092},
  year      = {2021},
  publisher = {IEEE},
  doi       = {10.1109/ICDE51399.2021.00205}
}

@inproceedings{fuchs2020edgeframe,
  author    = {Fuchs, Per and Boncz, Peter and Ghit, Bogdan},
  title     = {{EdgeFrame}: Worst-Case Optimal Joins for Graph-Pattern Matching in {Spark}},
  booktitle = {Proceedings of the 3rd Joint International Workshop on Graph Data Management Experiences \& Systems (GRADES) and Network Data Analytics (NDA)},
  pages     = {1--11},
  year      = {2020},
  publisher = {ACM},
  doi       = {10.1145/3398682.3399162}
}

@article{koutris2018algorithmic,
  author    = {Koutris, Paraschos and Salihoglu, Semih and Suciu, Dan},
  title     = {Algorithmic Aspects of Parallel Data Processing},
  journal   = {Foundations and Trends in Databases},
  volume    = {8},
  number    = {4},
  pages     = {239--370},
  year      = {2018},
  doi       = {10.1561/1900000055}
}

@inproceedings{ketsman2018distribution,
  author    = {Ketsman, Bas and Albarghouthi, Aws and Koutris, Paraschos},
  title     = {Distribution Policies for {Datalog}},
  booktitle = {21st International Conference on Database Theory (ICDT 2018)},
  series    = {LIPIcs},
  volume    = {98},
  pages     = {17:1--17:22},
  year      = {2018},
  doi       = {10.4230/LIPIcs.ICDT.2018.17}
}

@article{wang2015asynchronous,
  author    = {Wang, Jingjing and Balazinska, Magdalena and Halperin, Daniel},
  title     = {Asynchronous and Fault-Tolerant Recursive {Datalog} Evaluation in Shared-Nothing Engines},
  journal   = {Proc. VLDB Endow.},
  volume    = {8},
  number    = {12},
  pages     = {1542--1553},
  year      = {2015},
  doi       = {10.14778/2824032.2824052}
}

@misc{sun2026scaling,
  author        = {Sun, Yihao and Qi, Kunting and Gilray, Thomas and Kumar, Sidharth and Micinski, Kristopher},
  title         = {Scaling Worst-Case Optimal {Datalog} to {GPUs}},
  year          = {2026},
  eprint        = {2604.20073},
  archivePrefix = {arXiv},
  primaryClass  = {cs.DB},
  howpublished  = {arXiv preprint arXiv:2604.20073}
}

@article{zhu2017looking,
  author    = {Zhu, Jianqiao and Potti, Navneet and Saurabh, Saket and Patel, Jignesh M.},
  title     = {Looking Ahead Makes Query Plans Robust: Making the Initial Case with In-Memory Star Schema Data Warehouse Workloads},
  journal   = {Proc. VLDB Endow.},
  volume    = {10},
  number    = {8},
  pages     = {889--900},
  year      = {2017},
  doi       = {10.14778/3090163.3090167}
}

@article{yang2025accelerate,
  author    = {Yang, Yifei and Yu, Xiangyao},
  title     = {Accelerate Distributed Joins with Predicate Transfer},
  journal   = {Proc. ACM Manag. Data},
  volume    = {3},
  number    = {3},
  year      = {2025},
  doi       = {10.1145/3725259}
}

\end{document}